\documentclass[a4paper,fleqn,usenatbib]{mnras}

\usepackage{newtxtext,newtxmath}

\usepackage[T1]{fontenc}
\usepackage{ae,aecompl}

\usepackage{graphicx}	
\usepackage{amsmath}	
\usepackage{amssymb}	
\usepackage{subfig}

\title[Characterization of the Philae landing site]{Spectrophotometric characterization of the Philae landing site and surroundings with the ROSETTA/OSIRIS cameras}

\author[Hong Van Hoang et al.]{
Hong Van Hoang,$^{1,2}$\thanks{E-mail: Hong-Van.Hoang@obspm.fr}
S. Fornasier,$^{1,3}$
E. Quirico,$^{2}$
P. H. Hasselmann,$^{1}$
\newauthor M. A. Barucci,$^{1}$
H. Sierks,$^{4}$
C. Tubiana,$^{4}$
and C. G\"uttler,$^{4}$
\\
$^{1}$LESIA, Observatoire de Paris, Universit\'e PSL, CNRS, Universit\'e de Paris, Sorbonne Universit\'e, 5 place Jules Janssen,\\ 92195 Meudon, France\\
$^{2}$Universit\'e Grenoble Alpes, CNRS, Institut de Plan\'etologie et Astrophysique de Grenoble (IPAG), UMR 5274, Grenoble F-38041, France\\
$^{3}$Institut Universitaire de France (IUF), 1 rue Descartes, 75231 Paris Cedex 05, France\\
$^{4}$Max-Planck-Institut f\"ur Sonnensystemforschung, Justus-von-Liebig-Weg, 3, 37077, G\"ottingen, Germany
}

\date{Accepted XXX. Received YYY; in original form ZZZ}

\pubyear{2020}

\begin{document}
\label{firstpage}
\pagerange{\pageref{firstpage}--\pageref{lastpage}}
\maketitle

\begin{abstract}
We investigate Abydos, the final landing site of the Philae lander after its eventful landing from the Rosetta spacecraft on comet 67P/Churyumov-Gerasimenko on 12 November 2014. Over 1000 OSIRIS level 3B images were analysed, which cover the August 2014 to September 2016 timeframe, with spatial resolution ranging from 7.6 m/px to approximately 0.06 m/px. We found that the Abydos site is as dark as the global 67P nucleus and spectrally red, with an average albedo of 6.5\% at 649 nm and a spectral slope value of about 17\%/(100 nm) at 50$^\circ$ phase angle. Similar to the whole nucleus, the Abydos site also shows phase reddening but with lower coefficients than other regions of the comet which may imply a thinner cover of microscopically rough regolith compared to other areas. Seasonal variations, as already noticed for the whole nucleus, were also observed. We identified some potential morphological changes near the landing site implying a total mass loss of 4.7-7.0$\times$10$^5$ kg. Small spots ranging from 0.1 m$^2$ to 27 m$^2$ were observed close to Abydos before and after perihelion. Their estimated water ice abundance reaches 30-40\% locally, indicating fresh exposures of volatiles. Their lifetime ranges from a few hours up to three months for two pre-perihelion spots. The Abydos surroundings showed low level of cometary activity compared to other regions of the nucleus. Only a few jets are reported originating nearby Abydos, including a bright outburst that lasted for about one hour.  
\end{abstract}

\begin{keywords}
Comets: individual: 67P/Churyumov-Gerasimenko -- Methods: data analysis -- Methods:observational -- Techniques: photometric
\end{keywords}



\section{Introduction}

On 12 November 2014, at 08h35 UTC, the Philae lander began its historic descent towards the surface of comet 67P/Churyumov-Gerasimenko (67P). After nearly 7 hours, the lander made its first touchdown on the surface of the comet at 15h34 at a site named Agilkia. However, since the anchoring system did not work as expected, Philae bounced back and made contact with the surface of the comet twice before its final touchdown at 17h31, at a site called Abydos \citep[see][Fig. 3]{Biele2015}. Two phases of in-situ operations were planned to follow the landing, however only a modified version of the first short term phase, i.e. First Scientific Sequence (FSS) was implemented before the lander entered hibernation on 15 November 2014 at 00h08. In the following year, the ``awakened" Philae established several successful contacts with the Rosetta spacecraft between June and July 2015, the last of which took place on 9 July 2015 at 17h45 \citep{Ulamec2016}. The lander was finally unambiguously imaged by the OSIRIS/NAC instrument on 2 September 2016 \citep[see][Fig. 1]{Ulamec2017}, after four extensive search campaigns that involved instruments onboard both Rosetta and Philae \citep{ORourke2019}.

As the first instrument to conduct an in-situ analysis of a comet, Philae was able to provide a number of unique results. The immediate surroundings of the lander were captured in great details thanks to the CIVA and ROLIS cameras onboard Philae, the former revealed a diverse surface morphology that includes a blacklit block, networks of ubiquitous fractures of lengths from sub-cm to tens of cm and nearly 700 ``pebbles'' of sizes between 3.7 and 12.5 mm \citep{Bibring2015, Poulet2016}, whereas ROLIS showed a lumpy surface that includes both relatively smooth dark areas and more rough and clumpy bright areas with no large-scale colour variations \citep{Schroder2017}. 

Surface strength at Abydos was difficult to be evaluated, as none of the Philae instruments were fully able to penetrate the landing site. Several lower limits were derived from this lack of success: at least 4 MPa of local penetration resistance and 2 MPa of uniaxial compressive strength from MUPUS \citep{Spohn2015}, over 2 MPa based on CASSE \citep{Biele2015}. \citealt{Knapmeyer2018} combined data from both instruments and suggested a layered structure at Abydos, including a thin (i.e. a few cm) and hard top layer to account for the MUPUS deflection and a thicker layer (10 - 50 cm) with shear modulus 3.6 - 346 MPa, Young's modulus 7.2 - 980 MPa. On the other hand, using the lander Philae as an impact probe, \citealt{Heinisch2019} estimated that the overall surface compressive strength of the impact sites did not exceed $\sim$ 800 Pa: 399$\pm$393 Pa where Philae had a collision with the rim of the circular Hatmehit depression, 147$\pm$77 Pa at the site of the second touchdown and 8$\pm$7 - 73$\pm$70 Pa for several scratch marks found close to Philae.

This paper aims to characterize the final landing site of Philae as observed by the OSIRIS instrument, which had made observations of the Abydos site as well as its surroundings from when the Rosetta spacecraft first approached comet 67P in early August 2014 to the final days of the mission in late September 2016 (see Table \ref{tab:slope}). These observations allow us to analyse various aspects of the final landing site, i.e. morphology, reflectance, spectrophotometry and activity, as well as the evolution of these properties throughout the two years of OSIRIS operations. \phantom{\citep{El-Maarry2016}}

The article is organised as follows: Section 2 summarises the OSIRIS observations of Abydos and the analysis that was performed on these observations. Section 3 describes the terrains surroundings the landing site, while section 4 and section 5 present the analysis of the reflectance and the spectrophotometric properties of Abydos, respectively. Section 6 shows the bright patches of exposed volatiles that were found at close distances to Abydos in OSIRIS images and section 7 focuses on the few instances of cometary activity captured by OSIRIS near the final landing site of Philae. Finally, in section 8 we discuss our findings and we compare them to the published results on comet 67P.

\begin{figure}
	\includegraphics[width=\columnwidth]{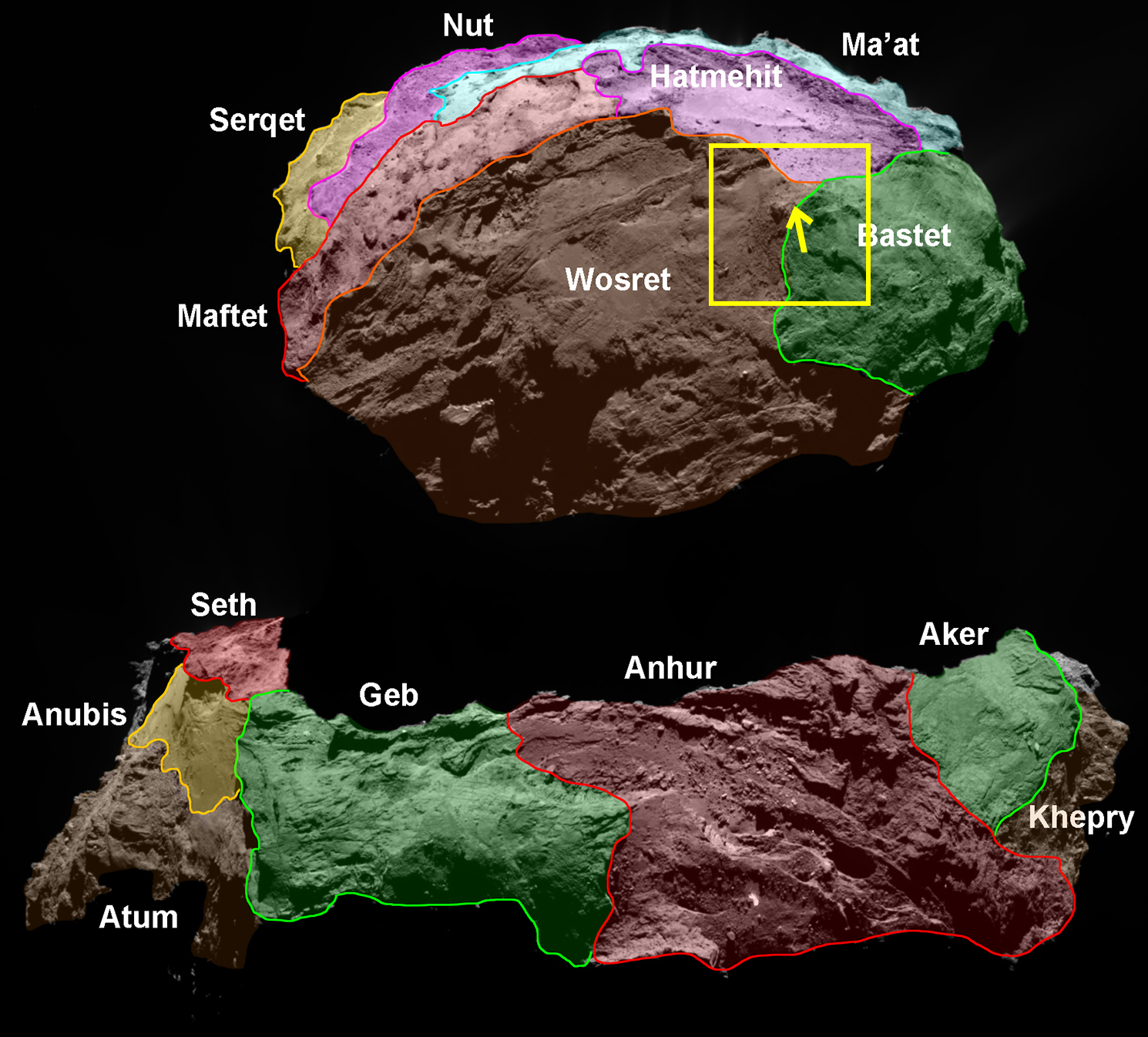}
    \caption{The southern hemisphere of the comet, superposed with regional boundaries. The arrow points to the Abydos site, while the box covers the 5$^\circ$ radius around Abydos. Regional boundaries refer to El-Maarry et al., A\&A 598, C2 (2017).}
    \label{fig:regions}
\end{figure}

\begin{figure}
	\includegraphics[width=\columnwidth]{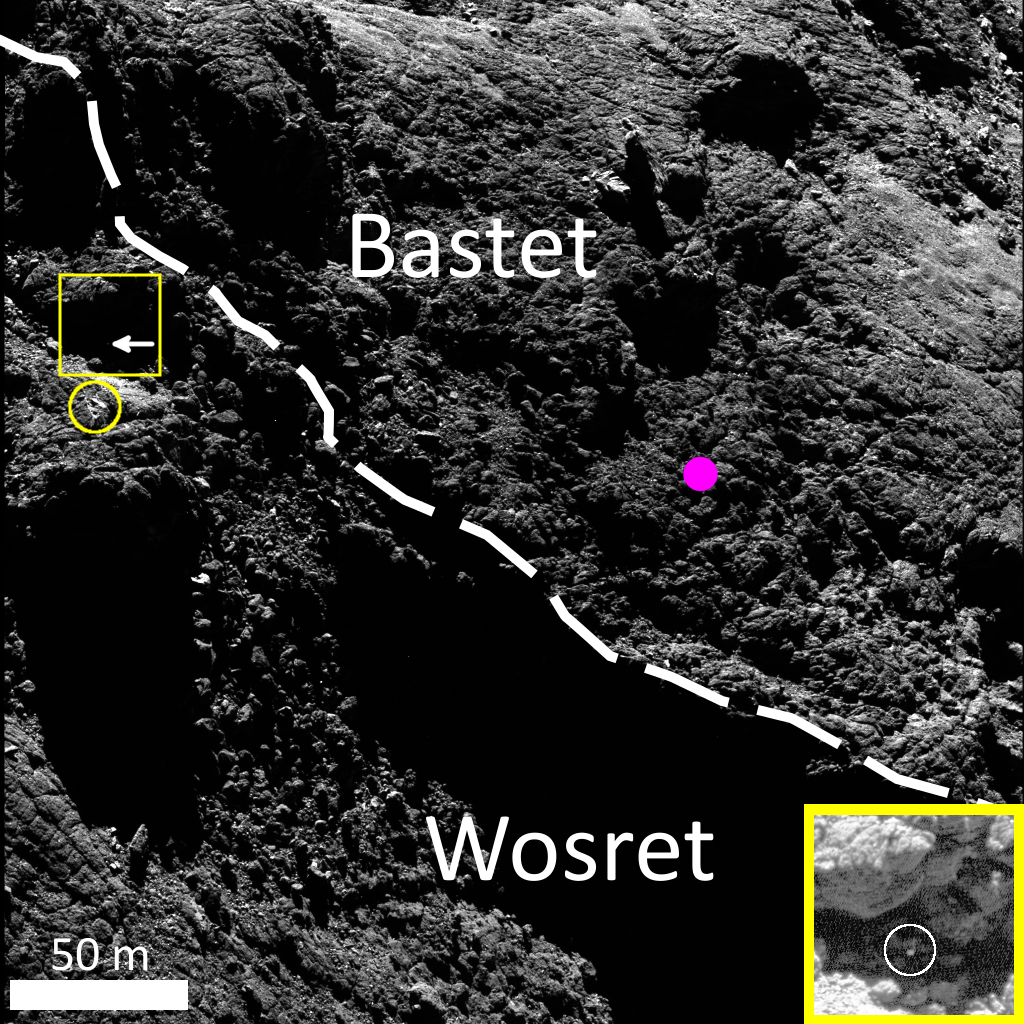}
    \caption{The Abydos (arrow) surroundings as captured by NAC on 1 June 2016, 9h12. The yellow circle indicates a bright feature that we named ``triangle'' near Abydos and the magenta dot represents the corresponding location of jet W-1 (see Table~\ref{tab:jets}). The image is superposed with a contrast-enhanced image of the area inside the box, from which the lander Philae can be perceived inside the white circle.}
    \label{fig:Abydos_stretch}
\end{figure}

\begin{table*}
\scriptsize
\caption{Observing conditions and the spectral slope of the surroundings of Abydos computed throughout the length of the Rosetta mission. The date of each sequence is the acquisition time of the reference orange filter (F22 or F82). Slope1 is the spectral slope and the associated error Std1 is evaluated in the 535-882 nm wavelength range. Slope2 and Std2 serve the same function in the 480-882 nm wavelength range.
	\newline $\Delta$: distance between the Rosetta spacecraft and the comet surface
	\newline $\alpha$: phase angle
	\newline N: Number of available filters. When N= 3, the three available filters are F22, F24 and F41, which required linear interpolation of the signals of the F24 and F22 filters in order to calculate the spectral slope in the 535-882 nm wavelength range.
	\newline [r5], [r10]: ROI is the smallest rectangle that covers a 5$^\circ$ and 10$^\circ$ radius from Abydos, respectively.
	\newline [Ab] Abydos is visible in the corresponding sequence.\
	}
\begin{tabular}{|p{3.7cm}|c|c|c|c|c|c|c|p{5.8cm}|}
\hline
\multicolumn{1}{|c|}{\textbf{Date}} & \textbf{$\Delta$ (km)} & \textbf{$\alpha$ ($^\circ$)} & \textbf{N} & \textbf{Slope1} & \textbf{Std1} & \textbf{Slope2} & \textbf{Std2} & \multicolumn{1}{c|}{\textbf{Description}} \\ \hline
2014-08-02T23:22:22$^\textrm{[r10]}$ & 395.3 & 27.1 & 7 & 15.1 & 2.1 & 16.4 & 2.0 & Mostly Hatmehit depression. \\ \hline
2014-08-03T20:40:22$^\textrm{[r10]}$ & 261.8 & 39.6 & 7 & 16.3 & 0.7 & 14.0 & 0.8 & Hatmehit and Bastet visible. \\ \hline
2014-08-06T00:20:22$^\textrm{[r5]}$ $^\textrm{[Ab]}$ & 118.9 & 49.6 & 7 & 17.4 & 1.6 & 19.6 & 1.3 & Mainly Hatmehit depression, with partially illuminated Wosret and Bastet \\ \hline
2014-09-02T07:44:02$^\textrm{[r5]}$ $^\textrm{[Ab]}$ & 51.5 & 40.7 & 9 & 17.6 & 2.3 & 18.8 & 1.2 & Mainly Hatmehit depression, but Bastet and (patially illuminated) Wosret also visible. F15 (269.3 nm) and F61 (931.9 nm) filters omitted from slope calculation. \\ \hline
2014-09-15T05:43:21$^\textrm{[r5]}$ $^\textrm{[Ab]}$ & 27.1 & 69.0 & 5 & 19.9 & 2.4 & 21.0 & 1.3 & Mainly Hatmehit depression, but Bastet and (patially illuminated) Wosret also visible. \\ \hline
2015-02-19T00:39:33$^\textrm{[r5]}$ $^\textrm{[Ab]}$ & 188.9 & 81.6 & 11 & 19.0 & 1.4 & 21.3 & 1.9 & Hatmehit, Wosret and Bastet all visible, but the latter two were poorly illuminated. \\ \hline
2015-02-21T12:54:04$^\textrm{[r5]}$ $^\textrm{[Ab]}$ & 69.5 & 44.3 & 3 & 15.5 & 1.3 & 17.7 & 1.5 & Hatmehit, Wosret and Bastet all visible. \\ \hline
2015-04-13T06:00:30$^\textrm{[r5]}$ $^\textrm{[Ab]}$ & 151.8 & 79.2 & 3 & 15.9 & 1.4 & 18.1 & 1.5 & Mostly Wosret and Bastet, but Hatmehit depression also visible. \\ \hline
2015-04-25T17:10:48$^\textrm{[r5]}$ & 91.5 & 65.3 & 11 & 17.2 & 1.6 & 19.5 & 1.5 & Mostly Bastet and Hatmehit, but Wosret also visible. \\ \hline
2015-05-02T10:43:57$^\textrm{[r5]}$ $^\textrm{[Ab]}$ & 123.9 & 61.2 & 11 & 15.4 & 1.1 & 17.0 & 1.1 & Mostly Wosret and Bastet, but Hatmehit depression also visible. \\ \hline
2015-05-16T22:56:24$^\textrm{[r5]}$ & 126.4 & 61.1 & 11 & 17.7 & 1.7 & 20.1 & 1.8 & Mostly Hatmehit rim and depression, but Bastet and Wosret also visible. \\ \hline
2015-05-22T15:35:55$^\textrm{[r5]}$ $^\textrm{[Ab]}$ & 129.0 & 61.1 & 10 & 15.3 & 1.3 & 17.9 & 1.1 & Mostly Wosret and Bastet, but Hatmehit also visible. F15 filter omitted from slope calculation \\ \hline
2015-06-18T13:00:04$^\textrm{[r5]}$ & 189.6 & 90.1 & 4 & 16.7 & 1.4 & 19.1 & 1.9 & Mostly Hatmehit rim and depression, but Bastet and Wosret also visible. \\ \hline
2015-06-27T07:15:18$^\textrm{[r5]}$ $^\textrm{[Ab]}$ & 198.0 & 90.0 & 11 & 16.7 & 1.0 & 18.5 & 1.3 & Mostly Wosret, but Bastet and Hatmehit also visible. \\ \hline
2015-07-04T13:43:26$^\textrm{[r5]}$ & 174.8 & 90.1 & 11 & 16.3 & 1.2 & 18.0 & 1.4 & Hatmehit, Wosret and Bastet all visible. \\ \hline
2015-07-11T18:07:20$^\textrm{[r5]}$ $^\textrm{[Ab]}$ & 157.2 & 89.6 & 11 & 16.5 & 0.8 & 18.4 & 0.8 & Mainly Wosret and Bastet. Hatmehit mostly in shadows. \\ \hline
2015-07-19T00:21:35$^\textrm{[r5]}$ & 179.3 & 89.6 & 11 & 16.1 & 1.0 & 17.6 & 1.1 & Hatmehit, Wosret and Bastet all visible. \\ \hline
2015-07-26T17:52:54$^\textrm{[r5]}$ $^\textrm{[Ab]}$ & 167.8 & 90.1 & 11 & 16.9 & 1.2 & 19.5 & 0.9 & Mostly Wosret, Bastet also visible. \\ \hline
2015-08-01T12:53:15$^\textrm{[r5]}$ $^\textrm{[Ab]}$ & 211.6 & 89.9 & 11 & 16.6 & 1.2 & 18.5 & 1.2 & Wosret and Bastet visible. \\ \hline
2015-08-09T17:43:25$^\textrm{[r10]}$ & 307.2 & 89.2 & 11 & 15.8 & 0.8 & 17.6 & 0.9 & Mostly Bastet and Hatmehit visible. Hatmehit depression partially in shadows. \\ \hline
2015-08-12T18:21:20$^\textrm{[r10]}$ & 328.6 & 89.6 & 11 & 16.0 & 1.1 & 17.7 & 1.1 & Mostly Bastet and Hatmehit visible. Hatmehit depression partially in shadows. \\ \hline
2015-08-22T23:18:04$^\textrm{[r10]}$ & 330.2 & 88.4 & 11 & 16.1 & 1.5 & 18.1 & 1.4 & Mainly Wosret, but Bastet and Hatmehit also visible. Hatmehit depression partially in shadows. \\ \hline
2015-08-30T23:55:56$^\textrm{[r10]}$ $^\textrm{[Ab]}$ & 403.1 & 70.2 & 11 & 15.8 & 1.1 & 17.7 & 0.9 & Wosret and Bastet visible. \\ \hline
2015-09-05T10:43:13$^\textrm{[r10]}$ & 427.9 & 100.5 & 7 & 16.1 & 1.1 & 18.1 & 1.1 & Wosret and Bastet visible, Hatmehit in shadows. \\ \hline
2015-10-11T22:15:56$^\textrm{[r10]}$ & 522.3 & 61.4 & 11 & 15.2 & 0.9 & 16.7 & 0.8 & Hatmehit, Wosret and Bastet all visible. \\ \hline
2015-10-20T01:03:29$^\textrm{[r10]}$ & 421.0 & 64.4 & 11 & 15.0 & 0.7 & 16.9 & 0.8 & Hatmehit, Wosret and Bastet all visible. A spot at (-7.3$^\circ$, -2.2$^\circ$), near the boundary with the shadows. \\ \hline
2015-10-31T19:09:07$^\textrm{[r10]}$ $^\textrm{[Ab]}$ & 293.3 & 62.1 & 11 & 15.5 & 0.6 & 16.9 & 0.7 & Wosret and Bastet visible. \\ \hline
\end{tabular}
\label{tab:slope}
\end{table*}

\begin{table*}
\scriptsize
\contcaption{Spectral slope of the surroundings of Abydos throughout the length of the Rosetta mission, continued from Table ~\ref{tab:slope}.
	\newline d - Spectral slope was calculated for only a small and flat area to reduce contribution from co-registration artefacts at high resolution. If the ROI covers two regions (Bastet and Wosret), we chose an area from the Wosret region.}
\begin{tabular}{|p{3.7cm}|c|c|c|c|c|c|c|p{5.8cm}|}
\hline
\multicolumn{1}{|c|}{\textbf{Date}} & \textbf{$\Delta$ (km)$^\textrm{a}$} & \textbf{$\alpha$ ($^\circ$)$^\textrm{b}$} & \textbf{N$^\textrm{c}$} & \textbf{Slope1} & \textbf{Std1} & \textbf{Slope2} & \textbf{Std2} & \multicolumn{1}{c|}{\textbf{Description}} \\ \hline
2015-11-19T20:08:20$^\textrm{[r10]}$ & 125.9 & 78.2 & 3 & 16.3 & 0.6 & 18.6 & 0.7 & Wosret and Bastet visible. \\ \hline
2015-11-22T11:42:23$^\textrm{[r5]}$ $^\textrm{[Ab]}$ & 128.6 & 89.6 & 3 & 17.2 & 1.2 & 19.7 & 1.5 & Wosret and Bastet visible. \\ \hline
2015-11-28T20:47:47$^\textrm{[r5]}$ $^\textrm{[Ab]}$ & 124.2 & 90.4 & 11 & 17.5 & 1.4 & 19.6 & 1.5 & Hatmehit, Wosret and Bastet all visible. Spectrally blue spots visible on the Hatmehit rim. \\ \hline
2015-12-07T01:14:32$^\textrm{[r5]}$ $^\textrm{[Ab]}$ & 97.9 & 89.7 & 3 & 17.0 & 0.8 & 19.7 & 0.9 & Hatmehit, Wosret and Bastet all visible, though the Hatmehit rim was mostly in shadows. Bright spots visible on the Hatmehit rim. \\ \hline
2015-12-10T01:32:27$^\textrm{[r5]}$ $^\textrm{[Ab]}$ & 101.6 & 89.8 & 3 & 17.4 & 1.0 & 20.0 & 1.1 & Mostly Wosret and Bastet, but Hatmehit depression also visible. \\ \hline
2015-12-26T16:06:26$^\textrm{[r5]}$ $^\textrm{[Ab]}$ & 76.7 & 90.2 & 11 & 16.8 & 1.8 & 19.5 & 1.9 & Hatmehit, Wosret and Bastet all visible. Parts of the Hatmehit rim that bordered Wosret were in shadows. Bright spots visible on the rim. \\ \hline
2016-01-09T16:06:26$^\textrm{[r5]}$ $^\textrm{[Ab]}$ & 77.8 & 90.5 & 11 & 18.7 & 1.6 & 20.9 & 1.5 & Mostly Hatmehit visible, although its rim was partially shadowed. Bright spots visible on the Hatmehit rim. \\ \hline
2016-01-17T05:00:45$^\textrm{[r5]}$ $^\textrm{[Ab]}$ & 84.0 & 63.1 & 3 & 16.1 & 0.7 & 18.4 & 0.9 & Hatmehit, Wosret and Bastet and visible, although the Hatmehit depression dominated the view. The Hatmehit rim was mainly in shadows, with bright spots visible. \\ \hline
2016-01-23T18:05:09$^\textrm{[r5]}$ $^\textrm{[Ab]}$ & 74.2 & 62.3 & 3 & 17.1 & 0.7 & 19.3 & 0.8 & Mostly Wosret, but Bastet and Hatmehit also visible. \\ \hline
2016-01-27T21:23:54$^\textrm{[r5]}$ $^\textrm{[Ab]}$ & 68.1 & 62.7 & 9 & 16.5 & 0.9 & 18.1 & 0.7 & Wosret and Bastet visible. \\ \hline
2016-02-10T19:21:45$^\textrm{[r5]}$ $^\textrm{[Ab]}$ & 47.3 & 65.2 & 11 & 17.8 & 1.1 & 19.5 & 1.1 & Mostly Wosret and Bastet, but Hatmehit depression also visible. \\ \hline
2016-02-13T09:00:20$^\textrm{[r5]}$ $^\textrm{[Ab]}$ & 44.0 & 69.6 & 3 & 17.7 & 1.2 & 20.2 & 1.2 & Mostly Wosret, but Bastet and Hatmehit also visible. \\ \hline
2016-04-09T17:35:16$^\textrm{[r5]}$ $^\textrm{[Ab]}$ & 32.7 & 40.8 & 11 & 16.0 & 1.0 & 18.4 & 1.2 & Hatmehit, Wosret and Bastet all visible. Spectrally blue spots visible on the Hatmehit rim. \\ \hline
2016-04-10T03:15:10$^\textrm{[r5]}$ $^\textrm{[Ab]}$ & 31.2 & 20.4 & 11 & 14.1 & 1.2 & 16.3 & 1.4 & Hatmehit, Wosret and Bastet all visible. \\ \hline
2016-05-16T22:24:52$^\textrm{d}$ $^\textrm{[r5]}$ $^\textrm{[Ab]}$ & 7.8 & 101.6 & 5 & 19.1 & 2.4 & 21.7 & 2.8 & Mostly Wosret visible. Frosts seen in some shadowed surfaces. \\ \hline
2016-05-28T12:28:24$^\textrm{d}$ $^\textrm{[r5]}$ & 4.9 & 101.0 & 5 & 19.7 & 2.1 & 22.7 & 2.4 & Only Wosret visible, which was dominated by shadows. \\ \hline
2016-06-12T22:29:58$^\textrm{[r5]}$ $^\textrm{[Ab]}$ & 27.5 & 81.4 & 3 & 18.3 & 2.0 & 21.0 & 2.2 & Wosret, Bastet and Hatmehit all visible. \\ \hline
2016-06-14T10:30:32$^\textrm{[r5]}$ $^\textrm{[Ab]}$ & 26.7 & 54.0 & 3 & 17.1 & 1.7 & 19.4 & 1.8 & Mostly Wosret and Bastet, but Hatmehit depression also visible. \\ \hline
2016-06-15T11:00:37$^\textrm{[r5]}$ & 26.8 & 44.9 & 3 & 16.9 & 1.1 & 19.1 & 1.2 & Mostly Hatmehit, but Bastet also visible. \\ \hline
2016-06-17T11:10:58$^\textrm{[r5]}$ & 30.1 & 62.1 & 3 & 17.2 & 1.5 & 19.8 & 1.7 & Hatmehit and Bastet visible. \\ \hline
2016-06-18T12:21:54$^\textrm{[r5]}$ & 30.5 & 79.4 & 3 & 18.1 & 1.6 & 20.7 & 1.9 & Only Hatmehit visible, with several bright spots on the rim. \\ \hline
2016-07-02T15:29:59$^\textrm{[r5]}$ $^\textrm{[Ab]}$ & 14.3 & 92.9 & 5 & 19.6 & 1.7 & 22.0 & 1.4 & Hatmehit, Wosret and Bastet all visible. \\ \hline
2016-07-03T12:45:15$^\textrm{[r5]}$ & 7.2 & 102.1 & 3 & 18.9 & 1.8 & 21.8 & 2.1 & Only Wosret visible, which was mostly in shadows. \\ \hline
2016-07-09T15:35:15$^\textrm{[r5]}$ & 12.1 & 99.0 & 7 & 19.3 & 1.8 & 21.4 & 1.8 & Only Wosret visible, with a number of bright spots under shadows of boulders. Frosts visible under a boulder. \\ \hline
2016-07-23T18:25:59$^\textrm{[r5]}$ & 8.4 & 100.3 & 3 & 18.5 & 1.6 & 21.5 & 1.7 & Only Hatmehit visible, with the rim mostly in shadows. \\ \hline
2016-08-21T17:59:03$^\textrm{d}$ $^\textrm{[r5]}$ & 4.2 & 89.7 & 3 & 18.1 & 2.5 & 21.1 & 2.0 & Wosret and Bastet visible, the former mostly in shadows. \\ \hline
2016-08-24T18:19:04$^\textrm{d}$ $^\textrm{[r5]}$ $^\textrm{[Ab]}$ & 3.9 & 91.5 & 3 & 18.6 & 1.8 & 21.9 & 2.0 & Wosret and Bastet visible, the former mostly in shadows. Weak frosts under some shadowed structures.  \\ \hline
\end{tabular}
\label{tab:slope_continued}
\end{table*}

\section{Observations and methodology}

Optical, Spectroscopic and Infrared Remote Imaging System (OSIRIS) was the name of the scientific imaging system onboard Rosetta. It was composed of two cameras: the Narrow Angle Camera (NAC, field of view $2.20^\circ \times 2.22^\circ$) that focused on the study of the nucleus composition and morphology, and the lower resolution Wide Angle Camera (WAC, field of view $11.35^\circ \times 12.11^\circ$), devoting to the study of the cometary coma. The NAC had 11 filters with a wavelength range of 250-1000 nm, while the WAC had 14 filters that covered the range of 240-720 nm \citep{Keller2007}. Our study is mostly based on NAC data, and every image of the comet surface used in this article originates from the NAC unless specified otherwise.

As the main instrument that was used in the search campaigns for Philae, the OSIRIS/NAC has taken over 1000 images of Abydos and its surroundings throughout the length of the Rosetta mission at various spatial resolutions that range from a few cm/px to more than 7 m/px. Lower resolution observations ($\ge$2 m/px) usually fell into a couple of phases of the mission: when the Rosetta spacecraft first approached the comet in August 2014 and around the perihelion passage, when the spacecraft was far from the comet because of the high activity; high resolution observations ($\le$ 20 cm/px) were mostly captured from May 2016 onwards. The landing site was observed at a wide range of phase angle from $\sim$20$^\circ$ to $\sim$120$^\circ$, though many of the observations were taken at high phase angles ($\ge\sim 90^\circ$) especially during the extended phase of the Rosetta mission in 2016. It must be noted that Abydos was frequently observed in poor illumination conditions (e.g. Fig.\ref{fig:Abydos_stretch} and Fig.~\ref{fig:Slope50}), making the characterization of the site particularly difficult.

The images used in our study have been corrected at OSIRIS level 3B from the OSIRIS pipeline, including corrections for bias, flat field and geometric distortions, calibration in absolute flux (W m$^{-2}$ nm$^{-1}$ sr$^{-1}$) and conversion to I/F radiance factor \citep{Tubiana2015, Fornasier2015}. Each image can be reconstructed using a stereophotoclinometric shape model (5 million or 12 million facets, \citealt{Jorda2016}), from which illumination conditions and observing geometry was retrieved for every pixel. NAIF SPICE kernels\footnote{\url{https://www.cosmos.esa.int/web/spice/spice-for-rosetta}} \citep{Acton2018} were sometimes used to obtain trajectorial and instrumental information relevant to the observing sequence.

\subsection{Spectrophotometry analysis}

\begin{figure}
\subfloat{\includegraphics[width=\columnwidth]{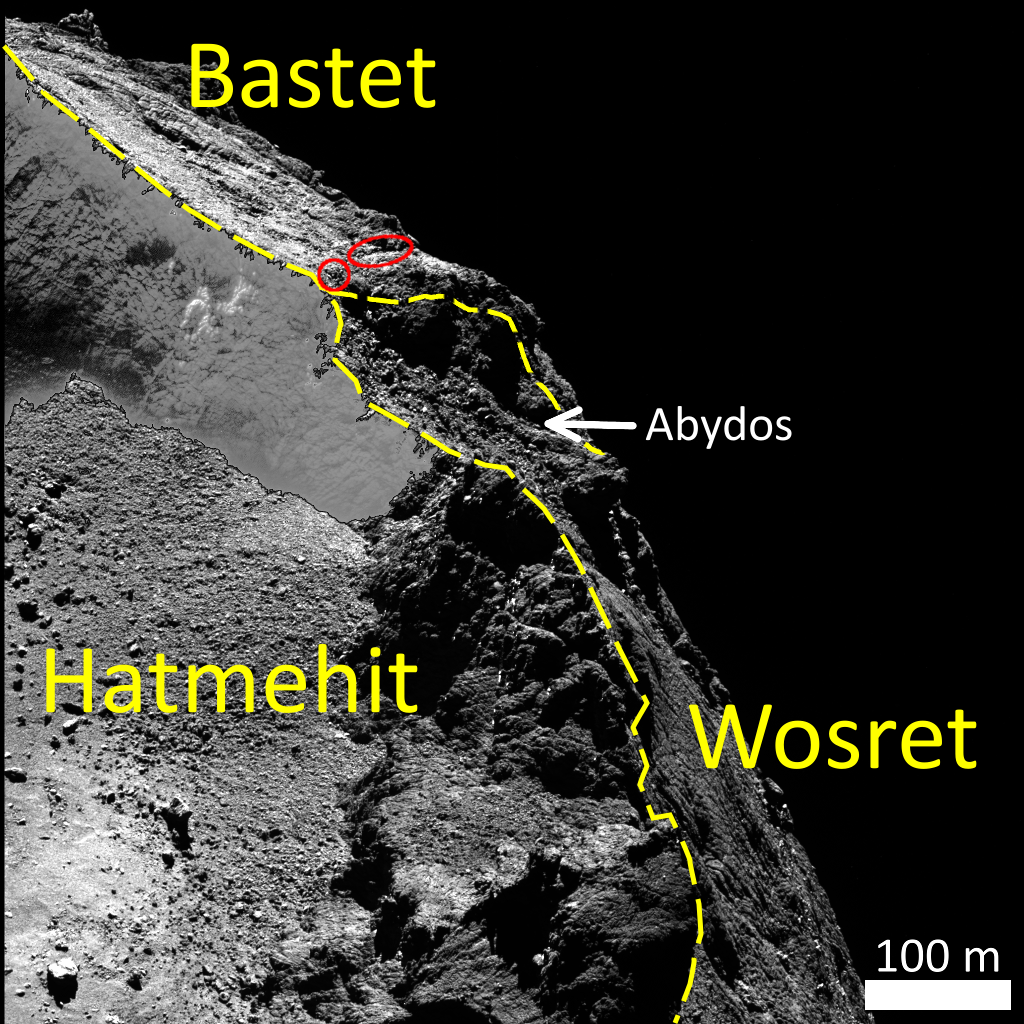}}
\caption{The comet nucleus as captured by the orange NAC filter on 5 March 2016, 10h35, superposed with a contrast-enhanced version of the dark side of the Hatmehit rim in the upper left side of the image. The arrow indicates the position of Abydos, and the red circle and ellipse respectively indicate the position of site AA and site BB (see Fig.~\ref{fig:Morphological_change_Abydos}), which will be discussed in Section 3.}
\label{fig:Hatmehit_rim}
\end{figure}

The spectral behaviour and evolution of the Abydos surroundings were investigated by the analysis of 54 colour sequences (i.e. when at least 3 different colour filters were available) that cover the time period from early August 2014 to late August 2016 (see Table \ref{tab:slope}). Firstly, a region of interest (ROI) covering an area of 5$^\circ$ (corresponding dimension $\sim$300 - 600 m) or 10$^\circ$ radius (corresponding ROI dimension $\sim$900 m)\footnote{The larger ROI was meant to compensate for the phases of low spatial resolutions, as mentioned previously in the second paragraph of Section 2.} was selected around the landing site. The coordinates of the ROIs follow the Cheops reference frame \citep{Preusker2015}.

All filters of a given sequence were first co-registered into a colour cube thanks to a python script based on the scikit-image library \citep{VanDerWalt2014}, and the optical flow algorithm \citep{Farneback2003}, in the same manner as described in \citealt{Fornasier2019a} and \citealt{Hasselmann2019}. The orange filter (either NAC/F22 or occasionally NAC/F82, both centred at 649.2 nm) of the sequence served as the reference, and one of those transformations was subsequently applied to the remaining filters: affine (including translation, rotation and shearing), similarity (including translation and rotation), projective and optical flow (which calculates the displacement field between two image frames). 

Once the filters were successfully stacked, false-colour RGB maps were created using the STIFF code \citep{Bertin2012} in order to make first visual inspections. Every RGB in this article follows this setting unless specified otherwise: ``green"= F22, ``blue"= F24 (480.7 nm) and ``red"= F41 (882.1 nm). Most of the comet surface appears ``red" in colour composites as it is dominated by a dark terrain, while bright volatile-rich patches are bright and white.

Each image was corrected for topographic-photometric conditions by applying the Lommel-Seelinger law\footnote{Pixels having incidence or emission angles above 80$^\circ$ were excluded}, which has been shown to work well on dark surfaces \citep{Li2015}. The reflectance of selected ROIs were calculated by integrating the flux in a box of 3$\times$3 pixels, and their relative reflectances were obtained by normalising the spectra to the green filter (NAC/F23 or occasionally NAC/F83, 535.7 nm) as commonly done in the literature for comet 67P in previous analyses (e.g. \citealt{Fornasier2015, Fornasier2019a, Deshapriya2016, Feller2016, Hasselmann2017}). If a sequence does not have a green filter, an artificial green filter was created through linear interpolation of the signals of the F24 and F22 filters. 

Finally, the spectral slope was calculated in the 535-882 nm range (expressed in \citealt{Fornasier2015}) and the 480-882 nm range. Spectral slope maps were generated using the F41 and F23/F83 filters. The average spectral slope was evaluated from a Gaussian fit of the slope distribution, in which the center of the peak represents the typical spectral slope value and the standard deviation of the fitted Gaussian acts as the error.

\section{Morphology}

\begin{figure}
\centering
\quad
    \subfloat[][19/01/2015, 08h04]{\includegraphics[width=0.9\columnwidth]{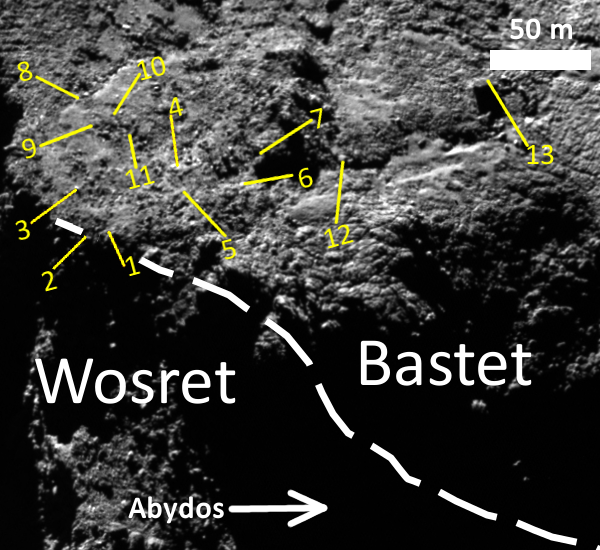}\label{fig:Morphological_change_Abydos_a}}
\newline
    \subfloat[][14/06/2016, 10h30]{\includegraphics[width=0.9\columnwidth]{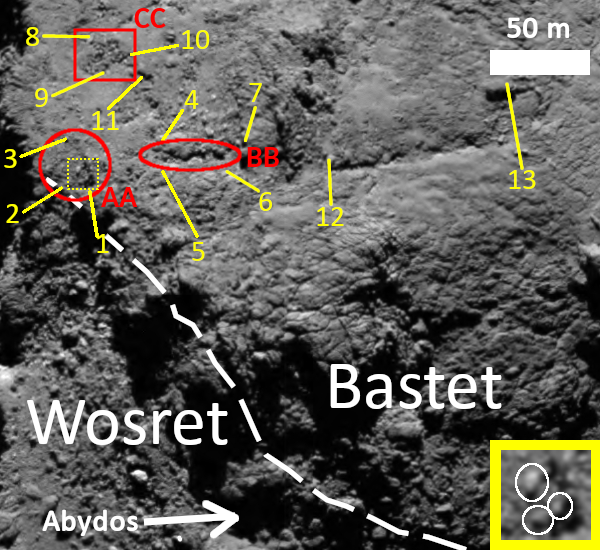}\label{fig:Morphological_change_Abydos_b}}
\caption{Parts of the Bastet region that is close to Abydos as imaged by the NAC/F22 on two separate dates: 19 January 2015 (top, pre-perihelion) and 14 June 2016 (bottom, post-perihelion). In the bottom image, the three sites of possible morphological changes AA, BB and CC are marked by red shapes, and the inset is a 3$\times$ magnification of the area within the yellow dotted box, where three boulders (within the white ellipses) can be seen within site AA. The top image has been rotated 15$^\circ$ to match the perspective of the bottom image. An animated version of this figure is provided in the online supplementary material.}
\label{fig:Morphological_change_Abydos}
\end{figure}

Abydos is located at -1.60$^\circ$ longitude and -8.04$^\circ$ latitude \citep{Ulamec2017}, which is in the Wosret region of the small lobe of the comet and very close to the boundary with the neighbouring Bastet region (see Fig.~\ref{fig:regions}). It was previously analysed in \citealt{Lucchetti2016} from pre-perihelion images. The landing site was located on an area covered in talus deposits that was surrounded by fractured and layered outcropping consolidated terrains. The Wosret and Bastet sides of the Abydos surroundings displayed similar fractured patterns, in which the fractures were typically 30-50 m long and roughly perpendicular to the Hatmehit rim. 447 boulders of size 0.8-11.5 m were identified near Abydos, and their distribution was characterized by a power-law index of -4.0$^{+0.3}_{-0.4}$ \citep[see][Fig. 2]{Lucchetti2016}. Similar distributions were found in several other regions of the comet e.g. the Nut region of the small lobe (-3.9$^{+0.3}_{-0.2}$), the Khepry/Ash boundary in the big lobe (-3.8$^{+0.1}_{-0.2}$) and the Seth/Ash boundary also in the ``body'' lobe (-4.2$^{+0.7}_{-1.1}$), which suggest that the possible boulder formation processes were gravitational events due to sublimation and headward erosion caused by thermal fracturing \citep{Pajola2015}.

High-resolution post-March 2016 images of the Abydos surroundings reveal two regions that were not previously featured in \citealt{Lucchetti2016}: a layered ``knobby'' terrain (the ``Bastet" side of Fig.~\ref{fig:Abydos_stretch}) and a second area of talus deposits under a prominent mound under Abydos, which is a frequently shadowed $\sim$9000 m$^2$ area covered in a number of small boulders of diameter below 10 m (as can be seen between the two major shadowed areas on the ``Wosret" side of Fig.~\ref{fig:Abydos_stretch} as well as in panel 12 of Fig.~\ref{fig:Abydos_RGB_spots} and the left side of Fig.~\ref{fig:Spectra_9Jul16}).

The Abydos site is also at close distances to a flat and circular depression of the Hatmehit region that covers 0.49 km$^2$ in area, 900 m in diameter and 150 m in depth \citep{LaForgia2015}. 175 boulders of size above 7 m have been identified from the depression floor, from which two power-law index values of distribution were derived: -3.4$^{+0.2}_{-0.1}$ (7-13 m) and -1.0$^{+0.1}_{-0.2}$ (13-22 m). The first index value is similar to the Abydos surroundings, whereas the second value suggests that the bigger boulders were the result of an evolution of the original material or were present in areas of high and continuous sublimation in the past or at present \citep{Pajola2015}. 

The Hatmehit region is separated from the Wosret and Bastet regions by consolidated terrains \citep{Giacomini2016}. While the Bastet/Hatmehit boundary (plus some of the Wosret/Hatmehit boundary) is a steep wall that shows layering at close inspection, the Wosret/Hatmehit boundary displays a complex terrace structure in high-resolution images taken in 2016: a big mound on top of the rim roughly 78 m long, 30 m wide and 48 m high; small, irregular ``steps'' from top to bottom and a roughly 10000 m$^2$ plateau (see Fig.~\ref{fig:Hatmehit_rim}).

Several differences between pre-perihelion and post-perihelion images of the Abydos surroundings have been found in the Bastet region (see the movie provided in the online supplementary material), however the lack of high-resolution observations and poor pre-perihelion illumination conditions prevented us from fully confirming that morphological changes had taken place. Possible sites of dust removal are located at the following coordinates\footnote{Order: (longitude, latitude).}: site AA, a V-shaped structure at (0.62$^\circ$, -4.70$^\circ$); site BB, a long fracture at (1.40$^\circ$, -5.44$.^\circ$) and site CC, a $\sim$288.9 m$^2$ shallow depression at (1.79$^\circ$, -3.92$^\circ$) (see Fig.~\ref{fig:Morphological_change_Abydos}). Since these features were often observed at high phase angles, their shadows could be used as a proxy for dimensions as expressed as follows:

\begin{equation}
h = L_\mathrm{shadow} \times \tan(\pi/2 - i)
\label{eq:shadow_length}
\end{equation}
where $h$, $L_\mathrm{shadow}$ and $i$ are respectively the landmark height, shadow length and incidence angle as estimated from the average of all the facets that intercept the tops and tips of the shadow. This method of measurement has been carried out for other regions of comet 67P \citep{El-Maarry2017, Hasselmann2019, Fornasier2019b} and other extraterrestrial bodies \citep{Arthur1974, Chappelow2002}.

A closer inspection revealed that the V-shaped structure composed of two ``walls'', one was 1.7$\pm$0.3 m high and 28.6$\pm$0.6 m long (between features 1 and 2 of Fig.~\ref{fig:Morphological_change_Abydos}) and the other was 2.1$\pm$0.3 high and 29.9$\pm$0.6 m long (between features 1 and 3 of Fig.~\ref{fig:Morphological_change_Abydos}), both of which enclosed three small boulders that were also not visible in pre-perihelion images (see the bottom inset of Fig.~\ref{fig:Morphological_change_Abydos}). The length and height of the long fracture were respectively 36.0$\pm$0.5 m and 4.1$\pm$0.1 m, and its width ranged between 2.0$\pm$0.5 at the ``waist'' near the middle to 4.0$\pm$0.5 m near both ends; while the depression had an irregular surface with depths below 0.8 m. If one applies a mean density of 537.8$\pm$0.7 kg/m$^3$ \citep{Preusker2017}, the approximate mass losses from the three sites were 0.9-1.9$\times$10$^5$ kg at site AA, 1.1-3.7$\times$10$^5$ kg at site BB and 0.16-1.2$\times$10$^5$ kg at site CC. These possible changes may have occurred during or close to the perihelion passage between June and October 2015 as the global cometary activity \citep{Vincent2016, Fornasier2019a} caused dust removal on the whole surface of comet 67P, leading to an overall ``bluer'' spectral slope (\citealt{Fornasier2016}, also discussed in Section 5), and their corresponding estimated mass losses are comparable with a few morphological changes reported in other regions on the surface of comet 67P: a cliff retreat with estimated mass loss of $0.8 \times 10^5$ kg in the Anhur region \citep{Fornasier2019b}; cavities C2 and C3 of the Khonsu region with estimated mass losses of $(4.1 \pm 2) \times 10^5$ kg and $(1.5 \pm 0.5) \times 10^5$ kg, respectively \citep{Hasselmann2019}. Other morphological changes reported by these two articles are linked to estimated mass losses ranging from a few to tens of million kilograms, which is 1-2 orders of magnitude higher than our estimations.
 
On the other hand, the area between boulders 12-13 and the nearby long fracture is a possible site of dust deposits, as the pre-perihelion fractured patterns appeared less pronounced post-perihelion. Another possible dust deposit site is located between site AA and CC (marked by features 3, 9 and 11 in Fig.~\ref{fig:Morphological_change_Abydos}), where the surface appears smoother post-perihelion than pre-perihelion (comparing the upper and lower panels of Fig.~\ref{fig:Morphological_change_Abydos}). Such dust may have been transferred from the three aforementioned sites of dust removal or nearby regions of the comet (e.g. Wosret, Sobek) when the surface dust fell back onto the surface of the comet as the level of cometary activity ceased with increasing distance from the Sun.

\section{Albedo}

\begin{figure}
\subfloat{\includegraphics[width=\columnwidth]{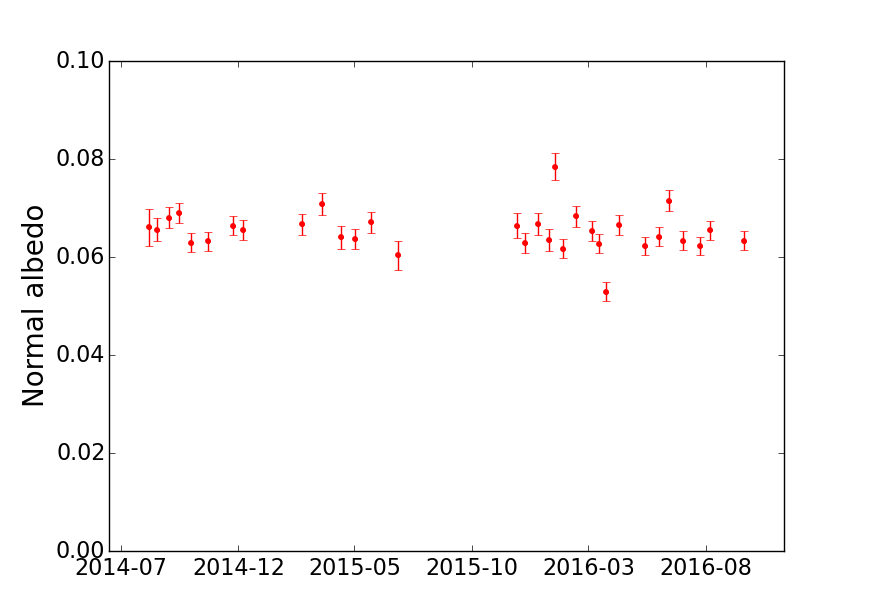}}
\caption{Evolution of the normal albedo of Abydos.}
\label{fig:albedo}
\end{figure}

\begin{figure}
\centering
	\includegraphics[width=0.6\columnwidth]{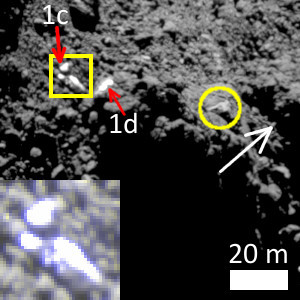}
	\caption{The area surrounding spots (1a) and (1b) as imaged by the NAC on 14 December 2014, 6h21, superposed with magnifications of the spots. (RGB setting: ``red"= ``green"= F22, ``blue"= F24). The arrow points to the position of Abydos, while the circle indicates the bright ``triangle'' mentioned in Section 4.}
	\label{fig:Spots_Sep14_evolution}
\end{figure}

\begin{figure*}
\centering
    \subfloat[][15/09/2014, 5h43 (3.4 AU)]{\includegraphics[width=0.27\textwidth]{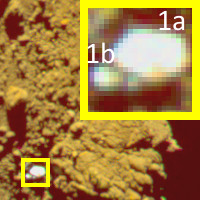}\label{fig:Abydos_RGB_spots_1}}
\quad
    \subfloat[][16/05/2015, 10h53 (1.6 AU)]{\includegraphics[width=0.27\textwidth]{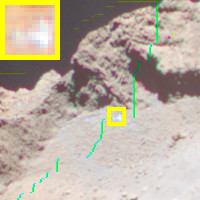}\label{fig:Abydos_RGB_spots_2}}
\quad
    \subfloat[][28/11/2015, 21h16 (1.8 AU)]{\includegraphics[width=0.27\textwidth]{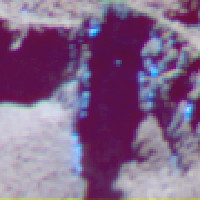}\label{fig:Abydos_RGB_spots_3}}
\newline
    \subfloat[][7/12/2015, 0h14 (1.8 AU)]{\includegraphics[width=0.27\textwidth]{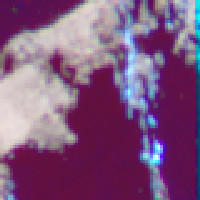}\label{fig:Abydos_RGB_spots_4}}
\quad
    \subfloat[][25/12/2015, 3h44 (2.0 AU)]{\includegraphics[width=0.27\textwidth]{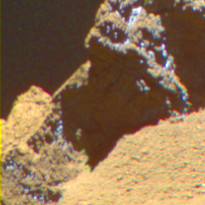}\label{fig:Abydos_RGB_spots_5}}
\quad
    \subfloat[][9/1/2016, 16h06 (2.1 AU)]{\includegraphics[width=0.27\textwidth]{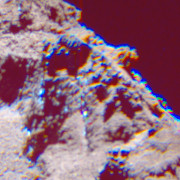}\label{fig:Abydos_RGB_spots_6}}
\newline
    \subfloat[][17/1/2016, 5h00 (2.1 AU)]{\includegraphics[width=0.27\textwidth]{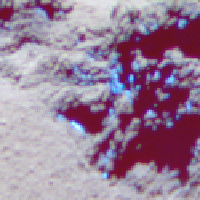}\label{fig:Abydos_RGB_spots_7}}
\quad
    \subfloat[][9/4/2016, 17h35 (2.7 AU)]{\includegraphics[width=0.27\textwidth]{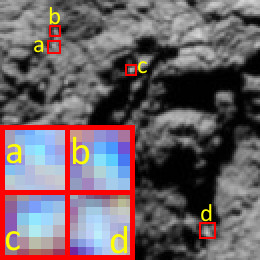}\label{fig:Abydos_RGB_spots_8}}
\quad
    \subfloat[][14/5/2016, 10h13 (3.0 AU)]{\includegraphics[width=0.27\textwidth]{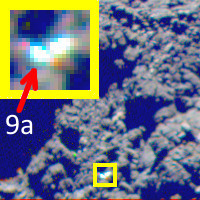}\label{fig:Abydos_RGB_spots_9}}
\newline
    \subfloat[][15/5/2016, 18h48 (3.0 AU)]{\includegraphics[width=0.27\textwidth]{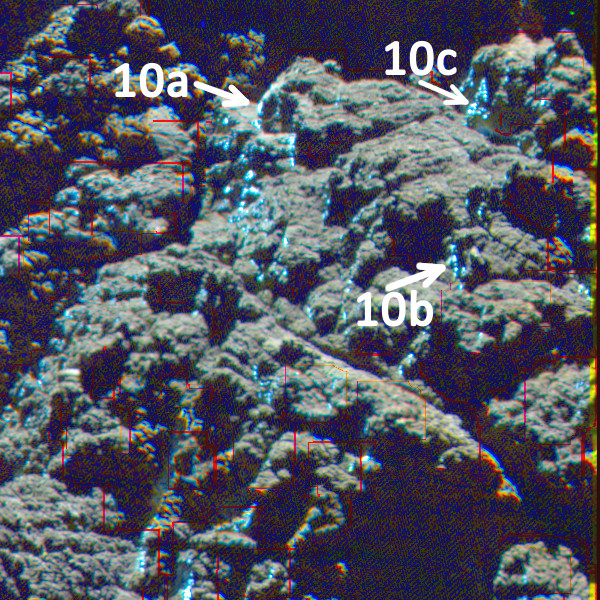}\label{fig:Abydos_RGB_spots_10}}
\quad
    \subfloat[][18/6/2016, 12h21 (3.2 AU)]{\includegraphics[width=0.27\textwidth]{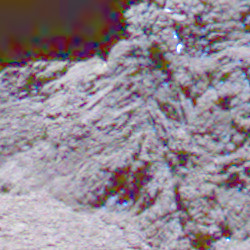}\label{fig:Abydos_RGB_spots_11}}
\quad
    \subfloat[][9/7/2016, 15h35 (3.4 AU)]{\includegraphics[width=0.27\textwidth]{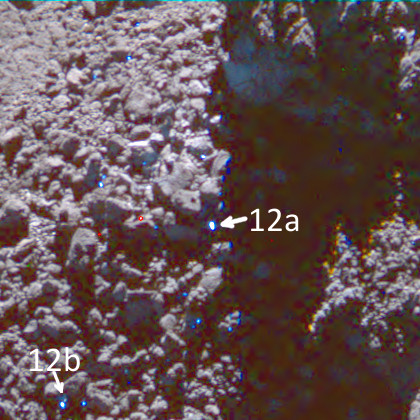}\label{fig:Abydos_RGB_spots_12}}
\newline
\caption{RGBs of bright spots found at close distances to Abydos. The RGB setting in panels 8 and 9 is ``red"= F22, ``blue"= F16 (360.0 nm) and ``green"= F24.}
\label{fig:Abydos_RGB_spots}
\end{figure*}

\begin{figure}
	\includegraphics[width=\columnwidth]{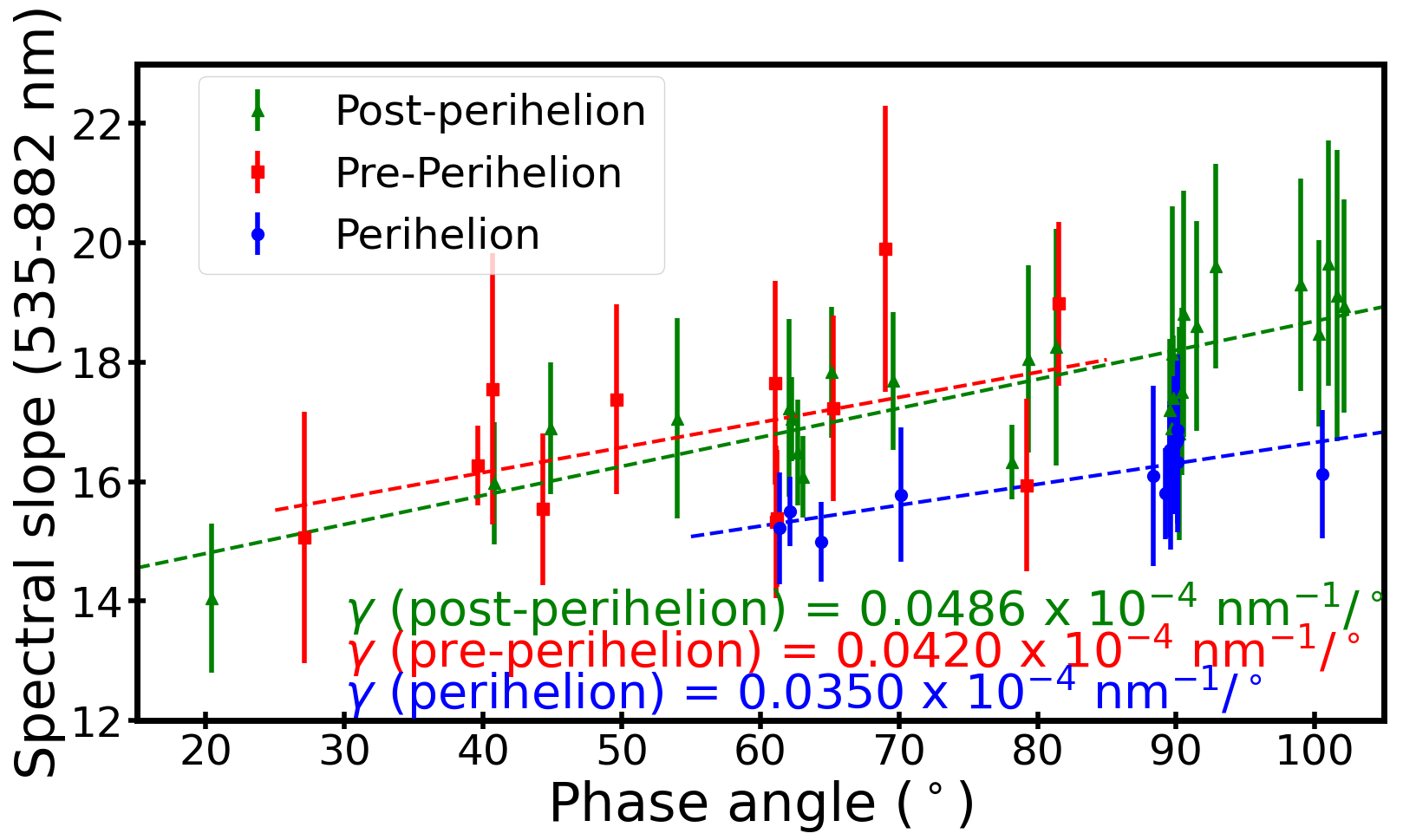}
    \caption{The spectral slope values of the Abydos surroundings in the 535.7-882.1 nm wavelength range versus phase angle from August 2014 to August 2016 observations. ``Perihelion'' is defined as the period between June and October 2015, when comet 67P was less than 1.6 AU from the Sun and reached perihelion on 13 August 2015, with a heliocentric distance of approximately 1.2 AU. The selected ROI in each case covers a 5$^\circ$ or 10$^\circ$ radius around Abydos, and more details can be found in Table~\ref{tab:slope}.}
    \label{fig:Slope}
\end{figure}

\begin{figure*}
\centering
    \subfloat[][15/09/2014, 5h43 ($\alpha= 69.0^\circ$)]{\includegraphics[width=0.3\textwidth]{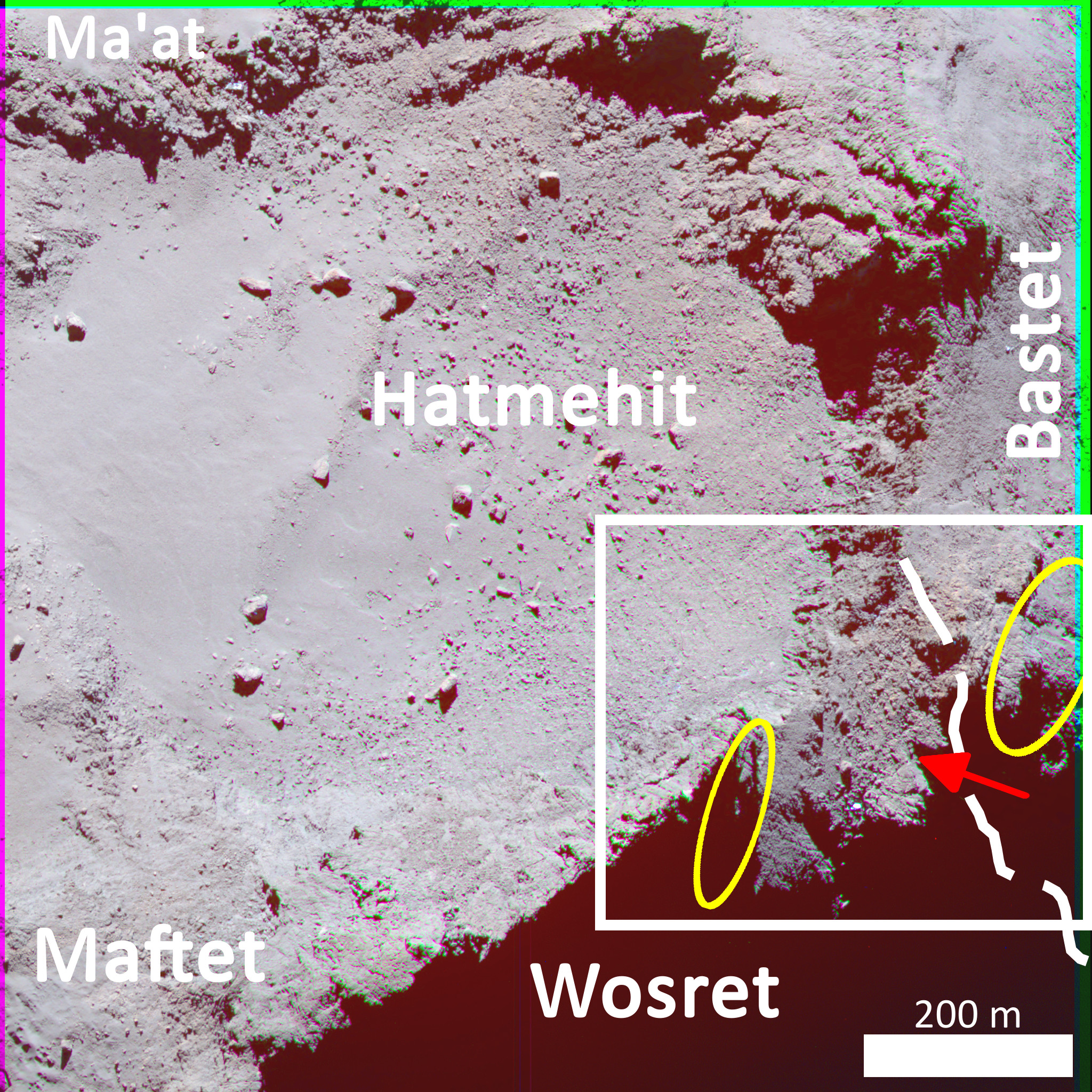}\label{fig:Slope70_a}}
\quad
    \subfloat[][30/08/2015, 11h43 ($\alpha= 70.2^\circ$)]{\includegraphics[width=0.3\textwidth]{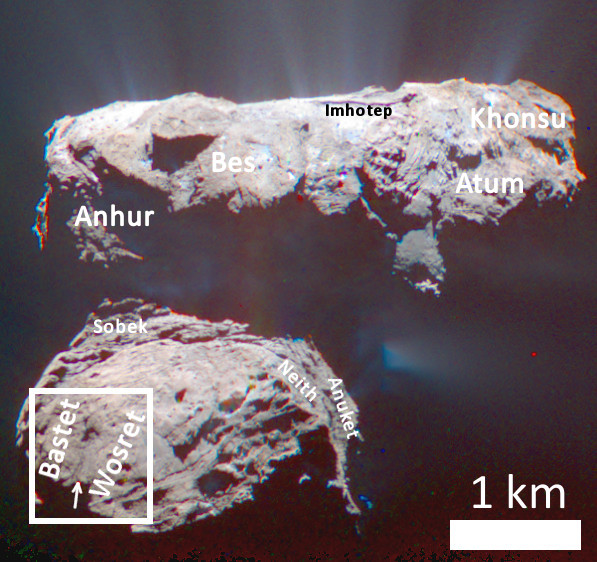}\label{fig:Slope70_b}}
\quad
    \subfloat[][19/02/2016, 19h21 ($\alpha= 65.2^\circ$)]{\includegraphics[width=0.3\textwidth]{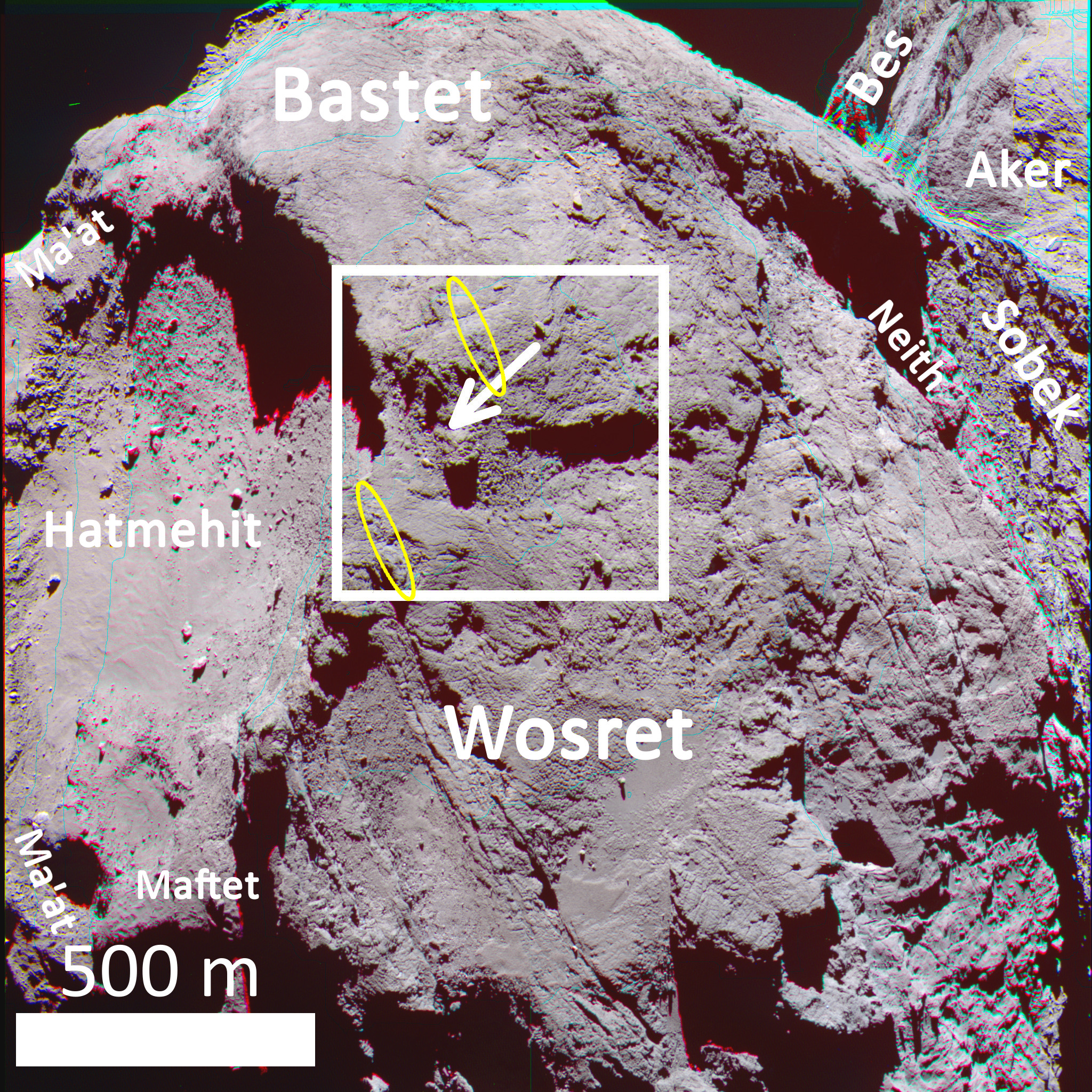}\label{fig:Slope70_c}}
\newline
    \subfloat[][slope= 19.9$\pm$2.4 \%/(100 nm)]{\includegraphics[width=0.3\textwidth]{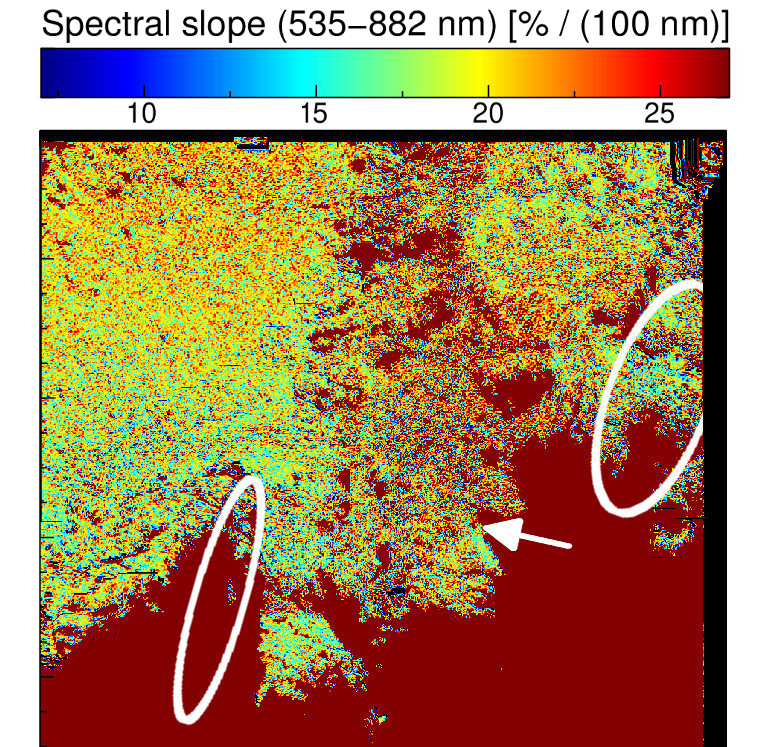}\label{fig:Slope70_d}}
\quad
    \subfloat[][slope= 15.7$\pm$0.9 \%/(100 nm)]{\includegraphics[width=0.3\textwidth]{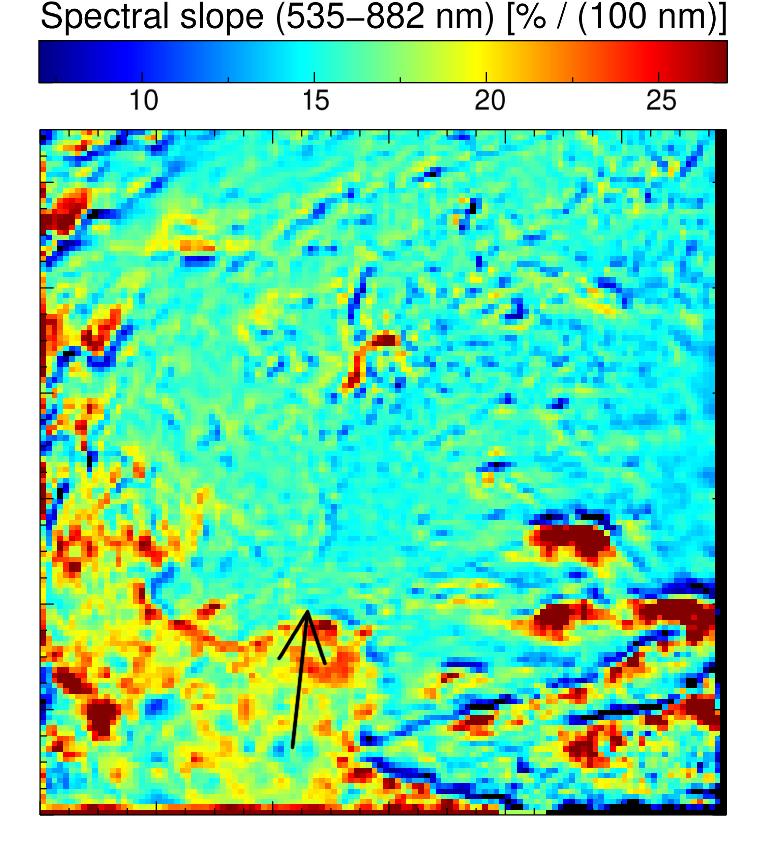}\label{fig:Slope70_e}}
\quad
    \subfloat[][slope= 17.8$\pm$1.1 \%/(100 nm)]{\includegraphics[width=0.3\textwidth]{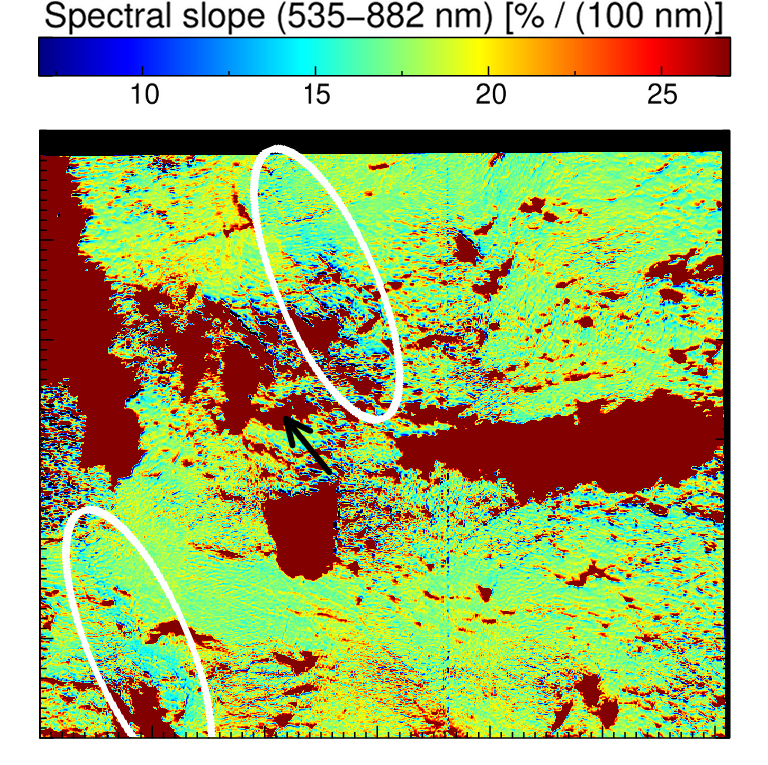}\label{fig:Slope70_f}}
\caption{\textit{Top}: The comet nucleus as imaged by NAC at phase angles $\alpha \sim 70^\circ$ on three different epochs: pre-perihelion, perihelion and post-perihelion, in which each box indicates the area around Abydos. \textit{Bottom}: Spectral slope maps (535-882 nm) of the area within the corresponding boxes, where reddest areas are often artefacts related to shadowed regions. Abydos is indicated by an arrow in every panel of this figure, and the ellipses cover the positions of the two spectrally blue ``stripes'' near the site.}
\label{fig:Slope70}
\end{figure*}

\begin{figure*}
\centering
    \subfloat{\includegraphics[width=0.4\textwidth]{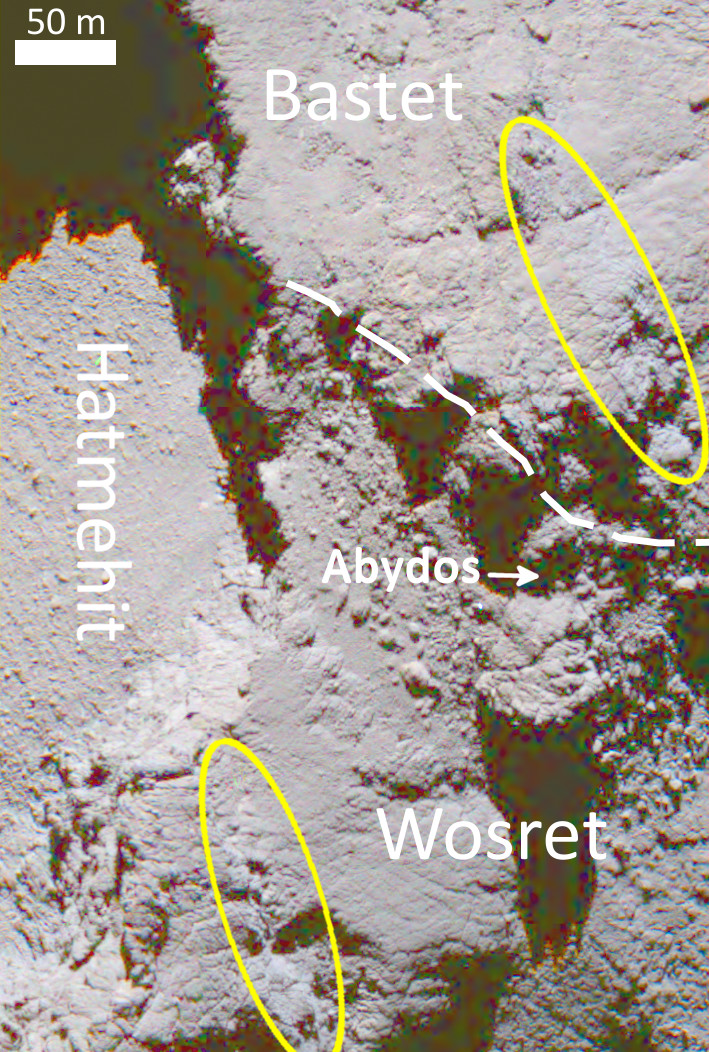}\label{fig:Slope50_a}}
\quad
    \subfloat{\includegraphics[width=0.45\textwidth]{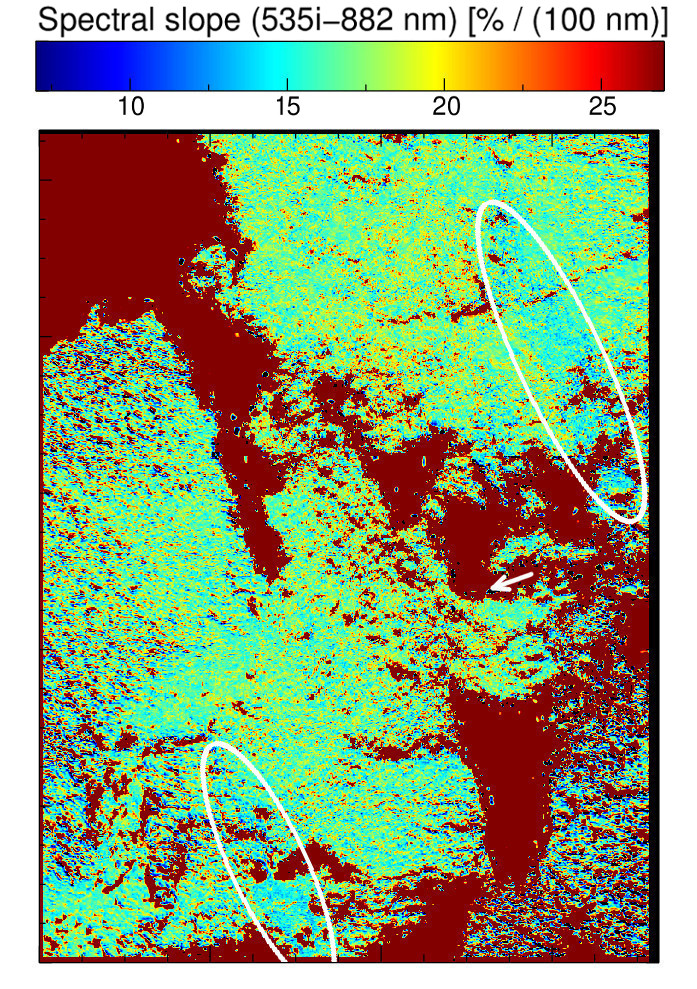}\label{fig:Slope50_b}}
\caption{Colour composite (left) and spectral slope in the 535-882 nm wavelength range (right) of the 5$^\circ$ area around Abydos (indicated by arrow) at 22h23, 14 June 2016, at a phase angle of 48.6$^\circ$. Gaussian fitted slope= 16.9$\pm$1.5 \%/(100 nm). The ellipses indicate the two spectrally blue ``stripes'' near Abydos.}
\label{fig:Slope50}
\end{figure*}

We studied the reflectance of Abydos by applying the \citealt{Hapke2008} model\footnote{Pixels having incidence or emission angles above 70$^\circ$ were omitted from our calculations.} using the parameters derived from the global analysis from \citealt{Hasselmann2017} (the ``all data'' set in Table 2): single-scattering albedo $w= 0.027$, shadow-hiding amplitude $B_0= 2.42$, shadow-hiding width $h_s= 0.081$, asymmetry factor $g_{sca}= -0.424$, average macroscopic roughness slope $\overline{\theta}= 26^\circ$ and porosity factor $K= 1.245$. Our dataset include 32 distinct observations at the orange filter, which covered two time periods: from when the Rosetta spacecraft first arrived at the comet in early August 2014 to June 2015, and from late November 2015 to near the end of the mission in late September 2016\footnote{Sequences taken near perihelion were omitted since they were taken at low spatial resolution, i.e. over 4 m/px.}. The ROI of each sequence covers a 0.5$^\circ$ radius around Abydos (corresponding dimension 40-60 m)\footnote{Unlike the spectral slope distributions which have a well-defined peak and can be Gaussian-fitted, the normal albedo distributions of many sequences are very irregular with large spread, hence a smaller ROI of 0.5$^\circ$ was chosen to reduce their effects on our calculations.}. We also attempted to generate a unique set of Hapke parameters for the Abydos site, however satisfactory results were not obtained due to the absence of observations near opposition and the unfavourable observing conditions of the Abydos site, which was often cast in shadows (e.g. Figures~\ref{fig:Abydos_stretch},~\ref{fig:Morphological_change_Abydos},~\ref{fig:Slope50}).

The average normal albedo of the dark terrain of the Abydos site is $\sim$6.5$\pm$0.2\% at $\lambda=$ 649.2 nm. This behavior is comparable to previous results reported for other regions of the comet at the same wavelength: 6.5\% for mostly the northern hemisphere of the comet in July-August 2014, 6.15\% for the area at the Imhotep/Ash boundary as observed by the NAC/F82 filter on 14 February 2015, 12:39:58 \citep{Feller2016} and 6.7\% at the original landing site Agilkia \citep{LaForgia2015}. The similarity in terms of brightness suggests that the dark terrains of the Abydos site have the same or very similar composition to the rest of the comet, from which several types of compounds have been suggested: dark refractory polycyclic aromatic hydrocarbons (PAHs) mixed with opaque minerals such as Fe-Ni alloys and FeS sulfides, carboxylic acids and NH$_4^+$ salts \citep{Quirico2016}. The reflectance of the dark terrains of the Abydos site appeared to stay relatively stable throughout the two years of OSIRIS observations. In fact no clear trend of evolution could be observed from the results (see Fig.~\ref{fig:albedo}), and there was no noticeable difference between pre-perihelion ($\sim$6.6$\pm$0.2\% on average) and post-perihelion values ($\sim$6.5$\pm$0.2\% on average).

Aside from the dark terrains, the Abydos surroundings also occasionally displayed highly localised bright areas\footnote{These bright spots were excluded from the calculation of the aforementioned reflectance of the dark terrains of Abydos.}. Most of them are relatively spectrally ``flat'' bright spots that indicate the presence of ices (e.g. H$_2$O or CO$_2$), such as the 0.26 m$^2$ spot (9a) as observed on 14 May 2016 (the arrow in the inset of panel 9 of Fig.~\ref{fig:Abydos_RGB_spots}), which will be further discussed in section 6. On the other hand, a few bright spots were spectrally ``red'' like the comet dark terrain, e.g. the ``triangle'' (as highlighted by the circles in Figures ~\ref{fig:Abydos_stretch} and ~\ref{fig:Spots_Sep14_evolution}) where the normal albedo reached up to $\sim$30-40\% and was visible from at least late November 2014 to late August 2016; or the 0.40 m$^2$ ``twin'' on the right of spot (9a), which was up to $\sim$10$\times$ brighter than the reference dark terrain and was also captured by NAC on 2 July 2016. The immediate surroundings of the Philae lander also displayed a similar dichotomy in brightness distribution as evidenced by data obtained from the two cameras onboard the lander: the ROLIS images show a combination of smooth dark areas and more rough and clumpy bright areas \citep{Schroder2017}, whereas the albedo of the features captured by CIVA ranged from 3-5\% as of the granular agglomerates with millimeter-to-centimeter grains to smoother textures with reflectance up to over 10\% \citep{Bibring2015}. The bright and spectrally red materials in OSIRIS images could be of similar nature to the bright spots observed by the Philae cameras, with possible causes including difference in texture or grain size \citep{Feller2016, Fornasier2017}, mineral grains and observing geometry that favours specular reflection \citep{Bibring2015}, the first two appear to be better explanations for long-lived features such as the ``triangle'' or the ``twin'' of spot (9a).

\begin{figure*}
	\includegraphics[width=\textwidth]{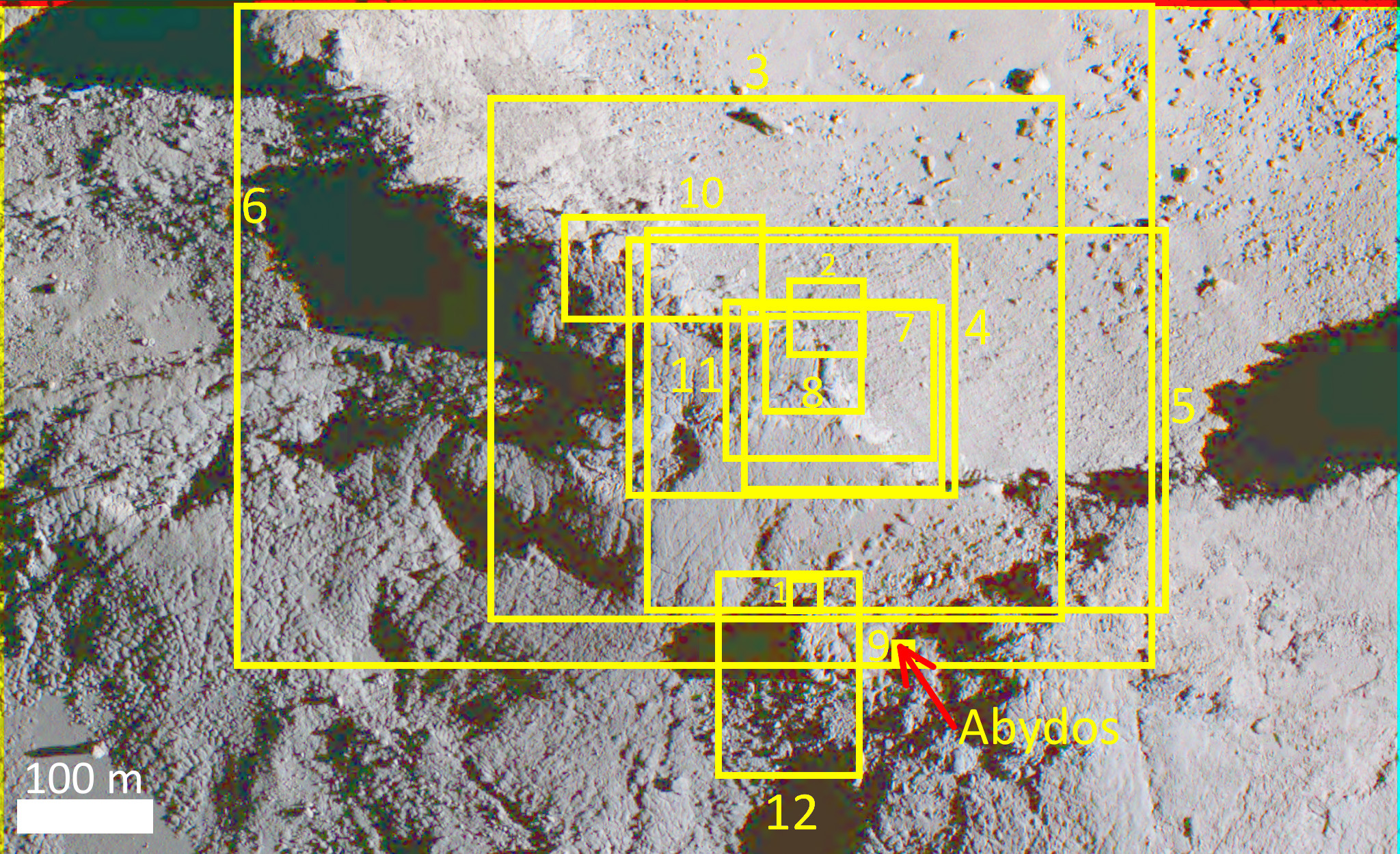}
    \caption{The positions of the spots in Fig.~\ref{fig:Abydos_RGB_spots} as superposed on the RGB map of the comet as captured by the NAC on 14 June 2016 at 10h30, with Abydos indicated by the red arrow. The area inside boxes 1, 2 and 9 represent the area inside the yellow boxes of the corresponding panel of Fig.~\ref{fig:Abydos_RGB_spots}. Note that the images shown in panels 2-7 and 10-11 were observed from the Hatmehit rim, which is a different perspective from this reference image.}
    \label{fig:Abydos-allspots}
\end{figure*}

\section{Spectral properties}

The 67P nucleus has been found to be heterogeneous in colours at several spatial scales. Three types of terrains are identified based on their spectral slope in the 535-882 nm range obtained at $\alpha \sim 50^\circ$: 10-14 \%/100 nm - relatively ``blue" terrains, 14-18 \%/100 nm - ``medium" terrains and 18-22 \%/100 nm - ``red" terrains \citep{Fornasier2015}. The first group is related to terrains somehow enriched in volatiles like Hapi or Seth while the last group has been linked to dust-covered regions. Abydos was observed at similar phase angles ($\sim$50$^\circ$) used by \citealt{Fornasier2015} in their analyses on two occasions: on 6 August 2014 at a resolution of 2.2 m/px, during which most of the Wosret and Bastet side of Abydos were under poor illumination conditions, and on 14 June 2016 at resolution 49.5 cm/px, when most of the Abydos surroundings were illuminated, albeit Abydos itself was in shadows (see Fig.~\ref{fig:Slope50}). The spectral slopes of Abydos fall between 16.5-17.5\%/(100 nm) in all the aforementioned observations, placing Abydos into the higher end of the medium terrain group.

Post-perihelion images reveal the presence of two ``stripes'' of moderately bright and blue terrains near Abydos: a $\sim$3800 m$^2$ stripe in the Bastet region that runs across a fracture and a couple of layers near the boundary with the Wosret region and a $\sim$3000 m$^2$ stripe in the Wosret region that covers a few strata (marked by pairs of ellipses in Figures~\ref{fig:Slope70} and~\ref{fig:Slope50}). The ``stripes'' were approximately 1.5-2.3$\times$ brighter than the reference dark terrain, and their spectral slopes at phase angles $\sim$50$^\circ$ (on 14 June 2016) were between 13.5-15.0 \%/100 nm. It is likely that these stripes already existed since at least September 2014 as the visible part of the Bastet ``stripe'' appeared spectrally bluer compared to the surroundings (roughly 15-17 \%/100 nm compared to 19.9$\pm$2.4 of the 5$^\circ$ radius around Abydos, see panels 1 and 4 of Fig.~\ref{fig:Slope70}), however poor illumination conditions in 2014 and the lack of pre-perihelion colour sequences at good spatial resolution complicate the detailed analysis.

From early August 2014 to late August 2016, OSIRIS/NAC has taken over 50 colour sequences of the Abydos surroundings (see Table~\ref{tab:slope}). These images were taken under a phase angle range of 20.4-101.6$^\circ$. Phase reddening i.e. the increase in spectral slope with increasing phase angle was observed throughout the aforementioned time period. During the perihelion passage, both the spectral slope values and the degree of phase reddening were lower compared to pre- and post-perihelion. Fig.~\ref{fig:Slope70} and Fig.~\ref{fig:Slope90} clearly show the spectral slope evolution of Abydos and its surroundings during different orbital periods: pre-perihelion in 2014, close to perihelion between June and October 2015, and post-perihelion in 2016. 

We analysed the spectral slope evolution over time for the complete data set. The spectral slopes evaluated in the 535-882 nm and 480-882 nm are reported in Table~\ref{tab:slope} for the individual observations together with a description of the regions covered by the observations. The spectral slope evolution is synthesised in Fig.~\ref{fig:Slope} (535-882 nm) and in Fig.~\ref{fig:Slope2441} (480-882 nm). The decrease of the spectral slope observed in Bastet, Wosret and Hatmehit was previously reported for the whole cometary nucleus by \citealt{Fornasier2016} and was attributed to the dust removal due to the intense cometary activity at close heliocentric distances \citep{Vincent2016, Fornasier2019a}. 

Phase reddening appeared to be more prominent after perihelion, as its corresponding linear coefficients is highest in both wavelength ranges, although the difference between the pre- and post-perihelion coefficients is less prominent in the case of the 480-882 nm spectral slope. The largest linear coefficents in the 535-882 nm wavelength range is $0.0486\pm0.00751 \times 10^{-4}$ nm$^{-1}/^\circ$, which is smaller than the value of $0.104 \times 10^{-4}$ nm$^{-1}/^\circ$ reported in \citealt{Fornasier2015}, covering the July and August 2014 observations of (mostly) the northern hemisphere of comet 67P at smaller phase angle values ($\alpha= 1.3-54.0^\circ$). Our coefficients are also smaller than the value $0.0652 \times 10^{-4}$ nm$^{-1}/^\circ$ for the area covered by the February 2015 flyby ($\alpha= 1.0-30.5^\circ$), which is located in the Imhotep region in the big lobe, close to the boundary with the neighbouring Ash region \citep{Feller2016}\footnote{The analysis of spectral slope in this article was conducted over a shorter wavelength range of 535-743 nm, normalized to the 535.7 nm wavelength.}. \citealt{Fornasier2016} reported a lower value of $0.041 \times 10^{-4}$ nm$^{-1}/^\circ$ for the observations between April and August 2015 at phase angles from 0 to 90$^\circ$, though it must be noted that this value is higher than the perihelion coefficient of $0.0350\pm0.00848 \times 10^{-4}$ nm$^{-1}/^\circ$ (June-October 2015, $\alpha= 61.4-100.5^\circ$) here reported for Abydos.

\section{Exposures of volatiles}

\begin{figure*}
\centering
    \subfloat{\includegraphics[width=0.2\textwidth]{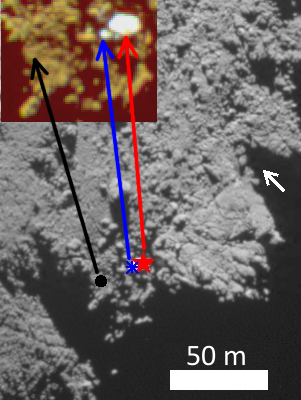}\label{fig:Spectra_twin_Sep14_1}}
\quad
    \subfloat{\includegraphics[width=0.35\textwidth]{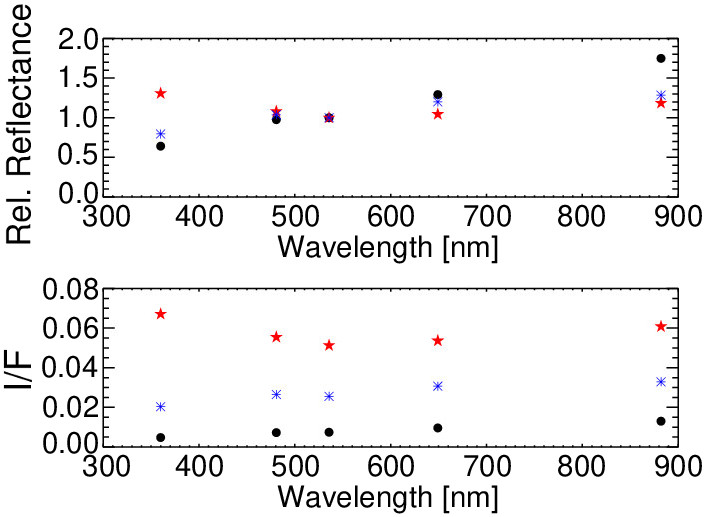}\label{fig:Spectra_twin_Sep14_2}}
\quad
    \subfloat{\includegraphics[width=0.35\textwidth]{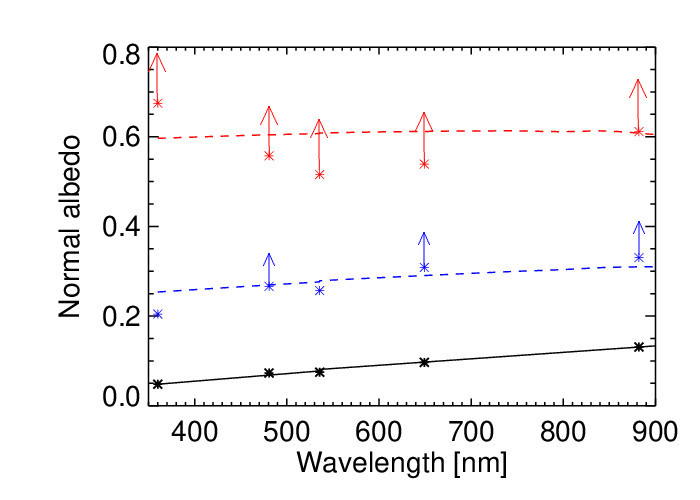}\label{fig:Spectra_twin_Sep14_3}}
\caption{\textit{Left}: The Abydos (white arrow) surroundings as imaged by the NAC/F22 on 15 September 2014, 5h43, superposed with an RGB of spots (1a) and (1b). \textit{Middle}: Spectra of spot (1a) (red star), (1b) (blue asterisk) and a reference dark terrain (black circle). \textit{Right}: The reflectances of the chosen points after correction to zero phase angle, with the arrows indicating the lower limit in normal albedo because the pixels were saturated. The black line is a linear fit of the dark terrain while the red and blue dotted lines are the best fits of the compositional model for 30$\mu$m ice grain size, indicating water ice abundances of $>\sim$57.6\% (red) and $>\sim$21.6\%.}
\label{fig:Spectra_twin_Sep14}
\end{figure*}

\begin{table*}
\scriptsize
\caption{A list of notable spots found near the final landing site of Philae, with S as the area of the spot and $\rho$ as the estimated water ice fraction.}
\begin{tabular}{|c|c|c|c|c|c|p{4.5cm}|}
\hline
\textbf{Lon ($^\circ$)} & \textbf{Lat ($^\circ$)} & \textbf{Obsevation time} & \textbf{S (m$^2$)} & \textbf{$\rho$ (30 $\mu$m)} & \textbf{$\rho$ (100 $\mu$m)} & \multicolumn{1}{c|}{\textbf{Description}} \\ \hline
-3.36 & -8.12 & 2 September 2014 - 14 December 2014 & 27.1 & $>\sim$57.6\% & $>\sim$60.6\% & Spot (1a), shown in panel 1 of Fig.~\ref{fig:Abydos_RGB_spots} \\ \hline
-3.50 & -8.09 & 2 September 2014 - 14 December 2014 & 3.4 & $>\sim$21.6\% & $>\sim$22.7\% & Spot (1b), shown in panel 1 of Fig.~\ref{fig:Abydos_RGB_spots} \\ \hline
-5.97 & -3.26 & 16 May 2015, 8h40 - 17 May 2015, 0h13 & 55.9 & 4.3\% & 4.4\% & Located in the Hatmehit depression, under the rim. Shown in panel 2 of Fig.~\ref{fig:Abydos_RGB_spots} \\ \hline
-1.48 & -8.10 & 14 May 2016, 10h10-10h14 & 0.26 & $<\sim$64.1\% & $<\sim$67.0\% & Spot (9a), shown in panel 9 of Fig.~\ref{fig:Abydos_RGB_spots}. About 5.4 m from Abydos, ``twin'' with a spectrally red bright spot. Poorly fitted models in all cases.  \\ \hline
-7.42 & -5.29 & 15 May 2016, 18h48-19h11 & 0.49 & 52.4\% & 54.7\% & Spot (10a), shown in panel 10 of Fig.~\ref{fig:Abydos_RGB_spots}. Poorly fitted models in all cases. \\ \hline
-8.53 & -5.38 & 15 May 2016, 18h48-19h11 & 0.19 & 29.1\% & 30.4\% & Spot (10b), shown in panel 10 of Fig.~\ref{fig:Abydos_RGB_spots}. Poorly fitted models in all cases. \\ \hline
-7.88 & -5.54 & 15 May 2016, 18h48-19h11 & 0.25 & 32.1\% & 33.6\% & Spot (10c), shown in panel 10 of Fig.~\ref{fig:Abydos_RGB_spots}. Poorly fitted models in all cases. \\ \hline
-5.55 & -4.54 & 15 May 2016, 18h48-19h11 & 0.15 & 17.8\% & 18.6\% & Spot (10d). \\ \hline
-5.79 & -4.78 & 17 June 2016, 11h31 - 18 June 2016, 12h22 & 8.9 & 39.7\% & 41.8\% & Spot (11a), shown in panel 11 of Fig.~\ref{fig:Abydos_RGB_spots}. \\ \hline
-2.72 & -9.45 & 9 July 2016, 15h35 - 16h05 & 1.2 & 80.6\% & 84.7\% & Spot (12a), shown in panel 12 of Fig.~\ref{fig:Abydos_RGB_spots}. Covered in shadows of nearby boulders and a prominent mound with frosts. \\ \hline
-3.07 & -10.25 & 9 July 2016, 15h05 - 15h36 & 0.6 & 29.3\% & 30.9\% & Spot (12b), shown in panel 12 of Fig.~\ref{fig:Abydos_RGB_spots}. Covered in shadows of nearby boulders. This spot appeared faint at its first observation at 15h03, but bright at the next observation at 15h33. \\ \hline
\end{tabular}
\label{tab:spots}
\end{table*}

A number of bright spots having a distinct flat spectrum compared to the comet average ``dark and red'' terrain was observed around the final landing site of Philae (see Fig.~\ref{fig:Abydos-allspots}). Previous joint studies with OSIRIS and VIRTIS have shown that regions having flat spectrophotometric behaviour also show water ice absorption bands in the NIR region \citep{Barucci2016, Filacchione2016}. We estimate water ice content of bright patches using geographical mixtures of volatiles and comet dark terrain (see Table~\ref{tab:spots}). In fact, the absence of water ice bands in the visible range prevents us from constraining relevant parameters necessary for more complex modelling (e.g. ice grain size). In the first step, normal albedo of the ROI was derived by correcting the illumination conditions and phase angle using the Hapke model parameters determined by \citealt{Hasselmann2017}. Then the water ice fraction $\rho$ of a bright spot was estimated by applying a linear mixture between the dark terrain and water ice:

\begin{equation}
	R_\mathrm{spot}= \rho R_\mathrm{ice} + (1-\rho)R_\mathrm{DT}
	\label{eq:Water_ice}
\end{equation}
where $R_\mathrm{spot}$ and $R_\mathrm{DT}$ are the measured reflectance of the spot and a reference dark terrain after applying the Hapke correction, respectively. We derived the water ice reflectance $R_\mathrm{ice}$ using the optical constant for grain sizes of 30 and 100 $\mu$m published in \citealt{Warren2008}. These values are adopted based on previous measurements of water ice grains on cometary nuclei \citep{Sunshine2006, Filacchione2016, Capaccioni2015}. We also attempted modelling with larger grain size (up to 3000 $\mu$m) but the fit of the spectral behaviour of the bright spots was poor.

Before perihelion, only a few patches were observed (e.g. panels 1-2 of Fig.~\ref{fig:Abydos_RGB_spots}, the best example being spot (1a) as one of the biggest\footnote{While the spot shown in panel 2 has a bigger estimated area, it was observed at relatively low spatial resolution of $\sim$2.4 m/px on 16-17 May 2015.}, brightest and longest-lived spots found at close distances to Abydos. It was first observed with its neighbour (1b) on 2 September 2014 at resolution $\sim$1.0 m/px and later at better spatial resolution of $\sim$0.5 m/px on 15-16 September 2014. In these observations, the two patches were so bright that pixels were saturated (see Fig.~\ref{fig:Spectra_twin_Sep14}). Estimation of the water ice content of spot (1a) yields a lower limit of 57.6-60.6\% for the two adopted grain sizes (see Fig.~\ref{fig:Spectra_twin_Sep14} and Table \ref{tab:spots}), which may indicate a very fresh exposure of ice. Similar bright spots covering the same period were already reported by \citealt{Pommerol2015} in different regions of the northern hemisphere of comet 67P.

The exposed patches of volatiles were observable up to December 2014, but the size of the spots progressively decreased to $\sim$4.6 m$^2$ (spot (1a)) and $\sim$1.6 m$^2$ (spot (1b)). Other spots appeared nearby in November and December 2014 images (see Fig.~\ref{fig:Spots_Sep14_evolution}): (1c) at (-3.5$^\circ$,-7.9$^\circ$), area $\sim$ 3.6 m$^2$; (1d) at (-3.2$^\circ$,-8.2$^\circ$), area 9.5 m$^2$. A possible factor in the long duration of spots (1a) and (1b) is the relatively low temperature of the Abydos surroundings in 2014: $\le$207 K at Abydos compared to the average 213$\pm$3 K for the dayside of the comet nucleus in August and September 2014 as recorded by VIRTIS \citep{Tosi2019} and 90-150 K during mid-November 2014 according to in-situ measurements \citep{Spohn2015, Lethuillier2016, Komle2017}.

\begin{figure*}
\centering
    \subfloat{\includegraphics[width=0.22\textwidth]{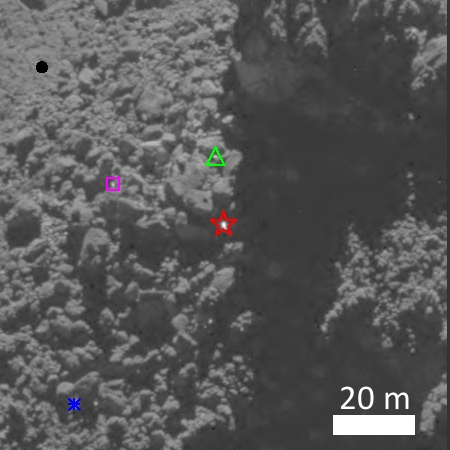}\label{fig:Spectra_9Jul16_1}}
\quad
    \subfloat{\includegraphics[width=0.32\textwidth]{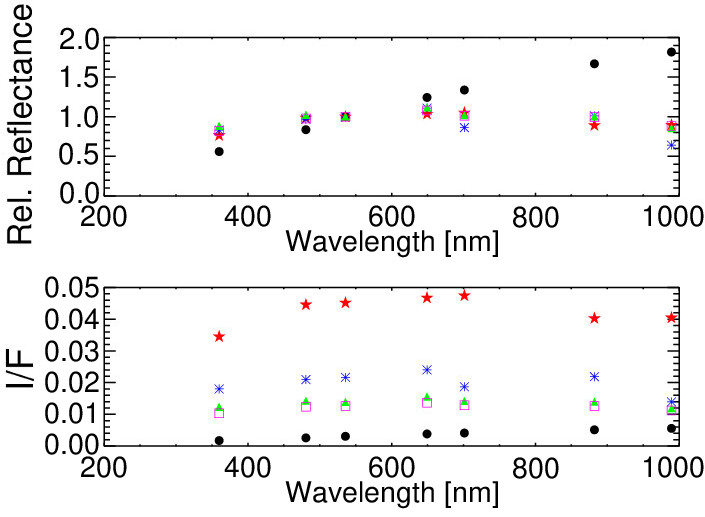}\label{fig:Spectra_9Jul16_2}}
\quad
    \subfloat{\includegraphics[width=0.32\textwidth]{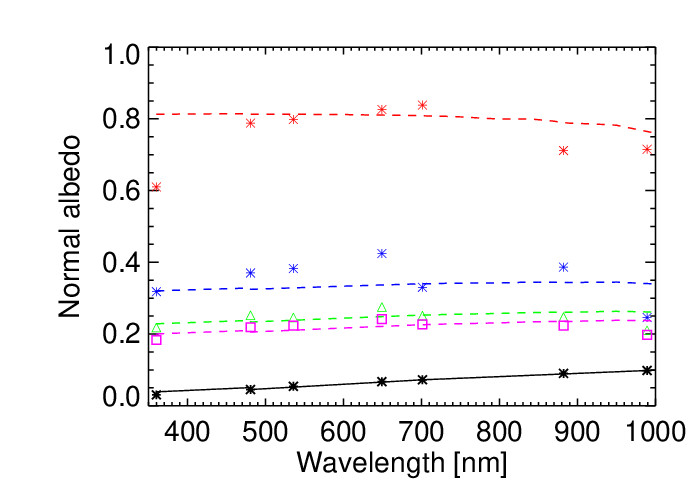}\label{fig:Spectra_9Jul16_3}}
\caption{\textit{Left}: Parts of the Abydos surroundings as captured by NAC/F22 on 9 July 2016, 15h35. The big mound on the top right corner is the same mound under Abydos as in Fig.~\ref{fig:Spectra_twin_Sep14}. \textit{Middle}: Spectra of the chosen points compared to a reference terrain (black), with spots (12a) and (12b) respectively represented by a red star and blue asterisk. The Lommel-Seelinger law was not applied for the bright spots. \textit{Right}: The black line is a linear fit of the dark terrain while the dotted lines are the best fits of the compositional model for 30$\mu$m ice grain size, indicating the water ice abundances of 80.6\% (red), 29.3\% (blue), 19.8 \% (green) and 16.8\% (magenta).}
\label{fig:Spectra_9Jul16}
\end{figure*}

Volatile exposures were imaged much more frequently after perihelion, with some areas repeatedly showing multiple bright patches in image sequences. A prominent case is the rim that separates the Hatmehit depression with the neighbouring Wosret region, which was shown to contain bright spots from late 2015 to near the end of the Rosetta mission. From late November 2015 to mid-January 2016, several colour sequences (e.g. panels 3-7 of Fig.~\ref{fig:Abydos_RGB_spots} captured multiple bright spots with area ranging between 5.5-35.0 m$^2$ \footnote{Note that these sequences have relatively low spatial resolution of 1.4-2.2 m/px.} and duration varying from $\ge\sim$30 minutes to at least one full day. The spots captured between late November 2015 and mid-December 2015 (e.g. panel 3-4 of Fig.~\ref{fig:Abydos_RGB_spots}) were relatively faint with estimated water ice percentage below 15\%, while some of the later spots (from late December 2015 to mid-January 2016, e.g. panel 5-6 of Fig.~\ref{fig:Abydos_RGB_spots}) were bright with estimated volatile fraction exceeding 50\% (see Fig.~\ref{fig:Spectra_Dec25}).

From March to May 2016, the Hatmehit rim was occasionally covered in tiny ($\le$1.0 m$^2$) bright materials under the shadows of its ``walls" structure, however colour sequences of these spots were not available until 15 May (panel 10 of Fig.~\ref{fig:Abydos_RGB_spots}). These spots lasted for at least half an hour, and compared to the reference dark terrain, these spots could be more than 10$\times$ brighter in the visible wavelength range and even brighter in the UV (NAC/F16, 360 nm), which corresponded to estimated local water ice abundance well above 50\%. These spots are very likely frosts and show the same spectral behaviour of frost in other regions \citep{Fornasier2016, Fornasier2019b}.

Some images captured during the final four months of the Rosetta mission feature bright patches on the Hatmehit rim (see Fig.~\ref{fig:spots_Hatmehit_late}), although the only colour sequence of such bright spots was taken on 18 June, 12h21. They were previously observed by the NAC/F22 the day before, spot (11a) being the largest with a 3.5 m$^2$ area and a fairly high estimated water ice fraction of $\sim$40\% (see Fig.~\ref{fig:Spectra_18Jun16}).

Another relatively ice-rich region at close distances to the final landing site was the area of talus deposits under the prominent mound near Abydos (see panel 12 of Fig.~\ref{fig:Abydos_RGB_spots}). High-resolution observations between May and July 2016 occasionally revealed tiny patches ($<$2 m$^2$) under the shadows of these boulders, and while that the site was observable up to the end of the Rosetta mission, the few available images only show that the area was dominated by shadows; however it is possible that ice-rich spots still existed in this area as exposed volatiles could be protected from solar illumination. The area was best observed on 9 July 2016 at a relatively high spatial resolution of 22.7 cm/px by 7 NAC colour filters (wavelength range 360-990 nm) as well as the WAC/F12 filter (629.8 nm)\footnote{The area was also observed at high resolution on two different dates: 14 May 2016 (15.3-16.2 cm/px) with only three NAC filters: F22, F24, F16 and on 24 July 2016 (16.3 cm/px) with only the NAC/F22 filter.}. On this day, several tiny bright spots were seen from this area with estimated water ice abundances ranging from 16\% to $\sim$80\%, which lasted for at least half an hour. The brightest spot with highest estimated ice fraction (represented by a red star and a red asterisk in Fig.~\ref{fig:Spectra_9Jul16}) appeared to be a fresh exposure of water ice, and it shows a negative slope in the NIR region that perhaps points to the presence of very large grains (i.e. a few mm) of ice or volatile species other than water or CO$_2$.

\section{Activity}

\begin{figure*}
\includegraphics[width= \textwidth]{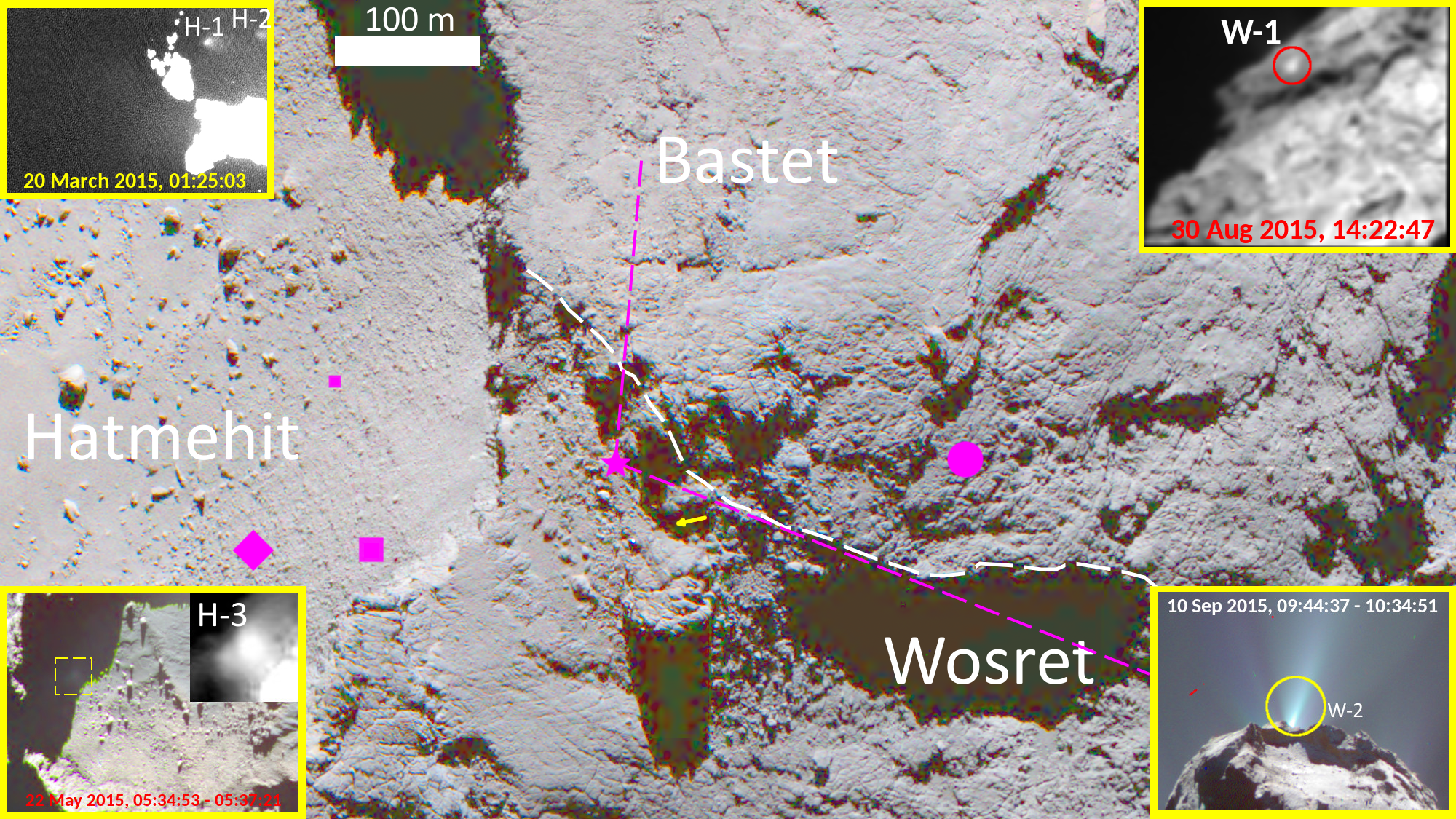}
\caption{The locations of the jets near Abydos (yellow arrow) as appeared on 14 June 2016, at 10h30, superposed with insets of their corresponding jets (RGB inset if possible). List of symbols: big square - H-1, small square - H-2, diamond - H-3, circle - W-1 and star - W-2. The white line indicates the Bastet/Wosret boundary, and the two magenta lines represent the uncertainty in finding the actual source of W-2.}
\label{fig:jets}
\end{figure*}

\begin{figure*}
\centering
    \subfloat{\includegraphics[width=0.25\textwidth]{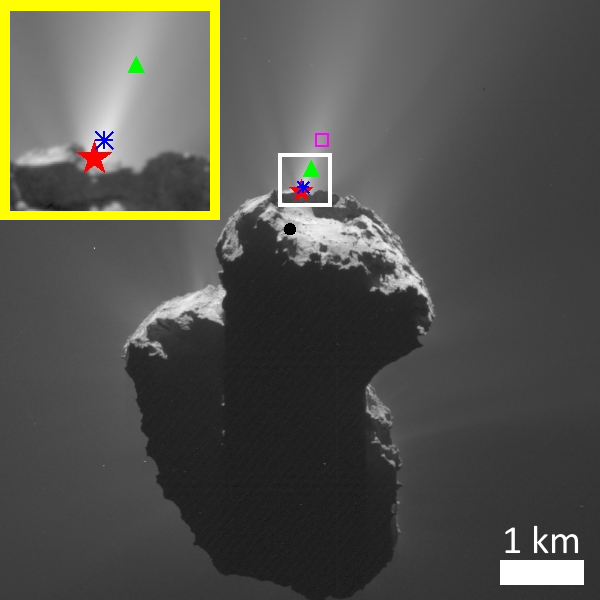}\label{fig:Spectra_W2_1}}
\quad
    \subfloat{\includegraphics[width=0.35\textwidth]{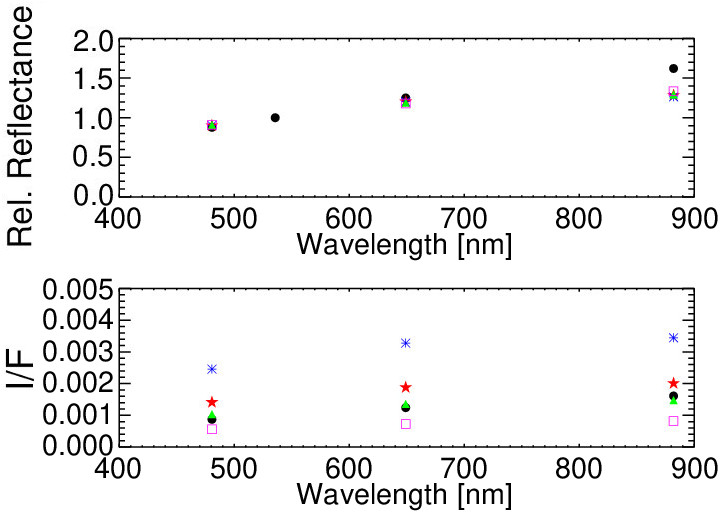}\label{fig:Spectra_W2_2}}
\quad
    \subfloat{\includegraphics[width=0.25\textwidth]{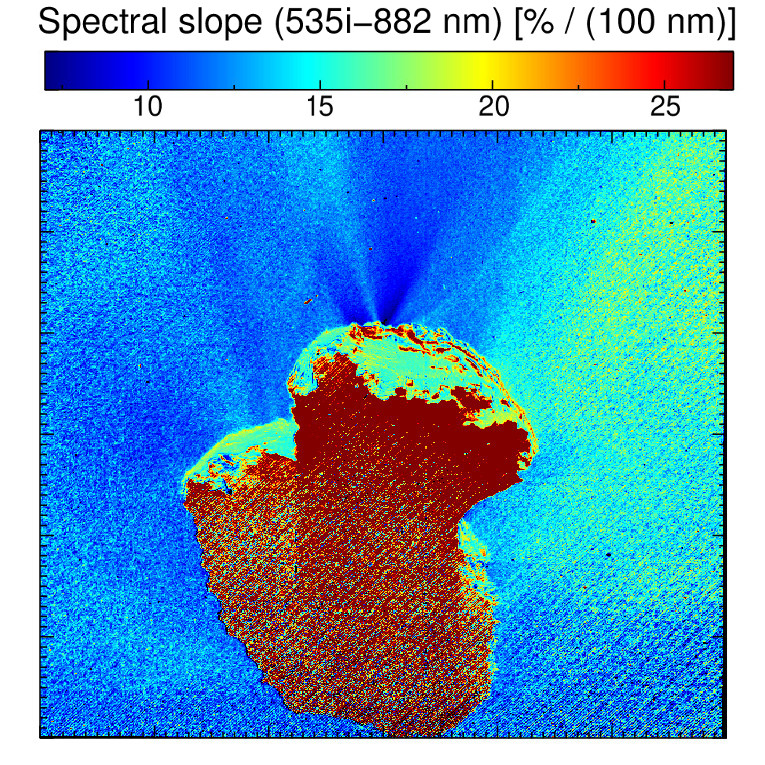}\label{fig:Spectra_W2_3}}
\caption{\textit{Left}: The comet nucleus as captured by NAC/F22 on 10 September 2015, 9h51, and the inset is a 2$\times$ magnification of the area inside the box. \textit{Middle}: Spectra of the chosen points of jet W-2 compared to a reference terrain (black). \textit{Right}: Spectral slope in the 535-882 nm range of the comet nucleus on 10 September 2015, 9h51, where reddest areas are often artefacts related to shadowed regions.}
\label{fig:Spectra_W2}
\end{figure*}

\begin{figure*}
\centering
    \subfloat{\includegraphics[width=0.25\textwidth]{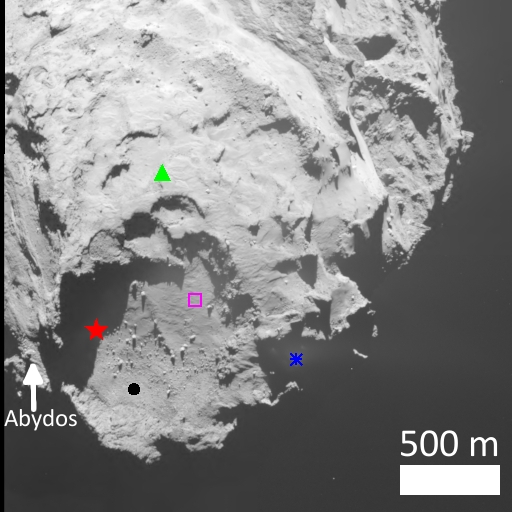}\label{fig:Spectra_H3_1}}
\quad
    \subfloat{\includegraphics[width=0.35\textwidth]{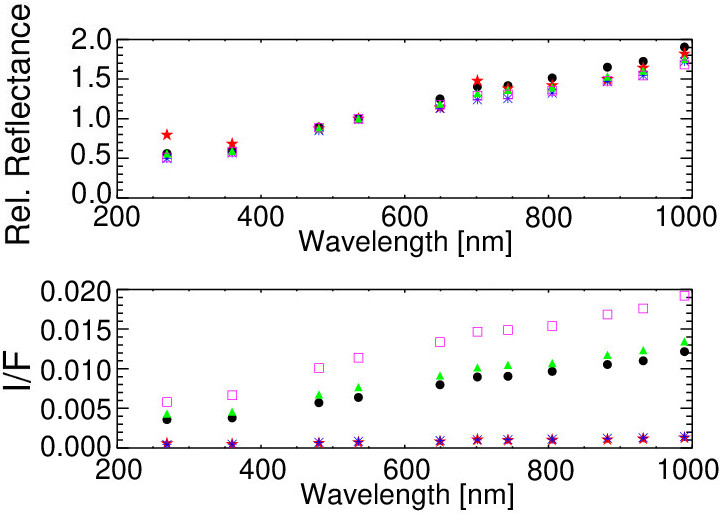}\label{fig:Spectra_H3_2}}
\quad
    \subfloat{\includegraphics[width=0.25\textwidth]{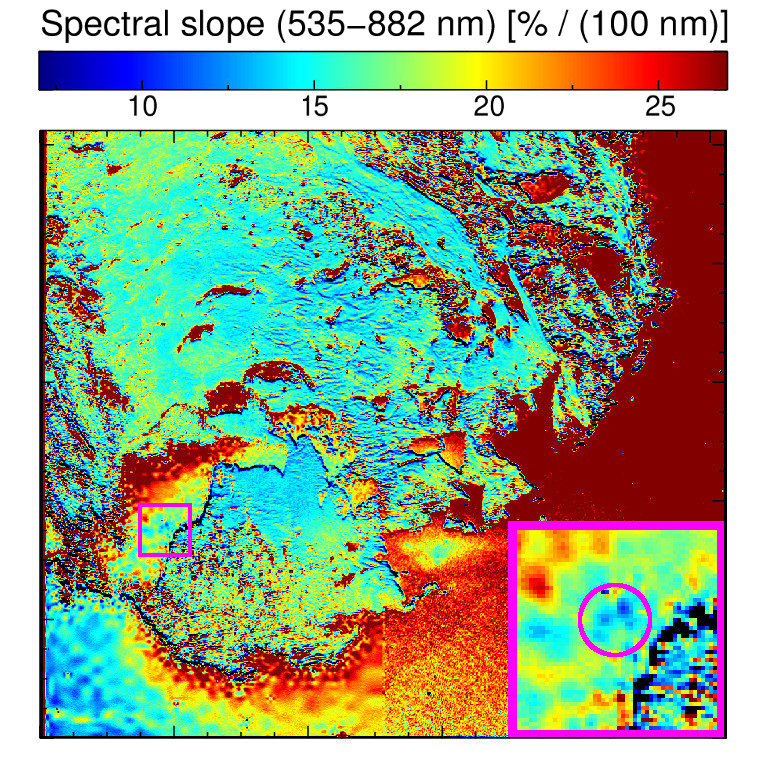}\label{fig:Spectra_H3_3}}
\caption{\textit{Left}: The comet nucleus as captured by NAC/F22 on 22 May 2015, 5h35. \textit{Middle}: Spectra of the chosen points compared to a reference terrain (black), where the red star indicates jet H-3 and the blue asterisk points to one of several faint jets from the Ma'at region. \textit{Right}: Spectral slope in the 535-882 nm range of the comet nucleus on 22 May 2015, 5h35, where reddest areas are often artefacts related to shadowed regions. The inset is a 2$\times$ magnification of the area inside the box, where jet H-3 is indicated by the circle.}
\label{fig:Spectra_H3}
\end{figure*}

\begin{table*}
\scriptsize
\caption{A list of jets found near Abydos, with $\alpha$ as the phase angle and D as the estimated diameter. The jets are classified in the same manner as in \citealt{Vincent2016}: A - collimated jet, B - broad plume and C - complex (broad$+$collimated). Note that the coordinates of the final jet was located on the limb at their time of observation, during which its actual source was obscured by the comet nucleus.}
\begin{tabular}{|c|c|c|c|c|c|c|p{5.5cm}|}
\hline
\textbf{Label} & \textbf{Lon ($^\circ$)} & \textbf{Lat ($^\circ$)} & \textbf{Observation time} & \textbf{$\alpha$ ($^\circ$)} & \textbf{Type} & \textbf{D (m)} & \multicolumn{1}{c|}{\textbf{Description}} \\ \hline
H-1 & -4.75 & -3.49 & 20 March 2015, 01:25:03 & 50.3 & B & $\sim$18 & A very faint jet from the Hatmehit depression, which was fully covered in shadows at the time. Only $\le \sim$2.2\% as bright as the dark terrains. \\ \hline
H-2 & -2.83 & -0.93 & 20 March 2015, 01:25:03 & 50.3 & B & $\sim$12 & A very faint jet from the Hatmehit depression, which was fully covered in shadows at the time. Only $\le \sim$1.8\% as bright as the dark terrains. \\ \hline
H-3 & -5.94 & -1.25 & 22 May 2015, 05:34:53 - 05:37:21 & 59.1 & B & $\sim$25 & A very faint and ``fuzzy'' jet from the shadowed part of the Hatmehit depression. Only $\le \sim$10-16\% as bright as the dark terrains. Relatively ``blue'' spectral slope of $\sim$14.4\%/ (100 nm) at a phase angle of 59.1$^\circ$. \\ \hline
W-1 & 2.08 & -10.78 & 30 August 2015, 14:22:47 & 70.2 & B & $\sim$30 & A jet that originated from a rough and fractured surface in the Bastet region, which was well-illuminated at the time of observation. \\ \hline
W-2 & -1.32 & -6.80 & 10 September 2015, 09:44:37 - 10:34:51 & 120.2 & C & $\sim$48 & Mini-outburst that was up to more than 2.5 times brighter than a reference dark terrain and has a relatively blue spectra in the NIR. \\ \hline
\end{tabular}
\label{tab:jets}
\end{table*}

No jets were directly observed from the Abydos site by the OSIRIS cameras, however the surroundings of the site did exhibit some level of cometary activity from March 2015 to September 2015 (see Fig.~\ref{fig:jets} and Table~\ref{tab:jets}). All three pre-perihelion jets found within a 5$^\circ$ radius of Abydos (i.e. H-1, H-2 and H-3) originated from the shadowed part of the Hatmehit depression. The first two jets were shown to be less than 3\% as bright as the reference terrain in the orange NAC/F22 filter. Jet H-3 offers some clues about the composition of such faint jets as it was observed during its $\ge$2 minutes duration by a sequence of 11 NAC colour filters on 22 May 2015, which showed that the jet was only $\le \sim$10-16\% as bright and spectrally similar to the dark terrain (see Fig.~\ref{fig:Spectra_H3}).

The Wosret/Bastet side of the Abydos surroundings appeared to begin its activity about two weeks after perihelion, which coincided with the peak of activity of the whole comet nucleus \citep{Fornasier2019a}. The earliest jet from this area is jet W-1 (see the top right corner of Fig.~\ref{fig:jets}), which was $\sim$1.4$\times$ brighter than its immediate surroundings as captured by the NAC/F15 on 30 August 2015. Peak of the activity near Abydos seemed to occur near mid-September 2015 as a mini-outburst was seen by OSIRIS/NAC on 10 September (W-2), which lasted for nearly one hour and was up to 2.5 times brighter than the dark terrain. The mini-outburst displayed a relatively ``bluer'' spectra in the VIS+NIR region compared to the dark surface of the comet (see Fig.~\ref{fig:Spectra_W2}), e.g. the spectral slope in the 535-882 nm of the sampling points of jet H-3 was 7.6-9.7\%/(100 nm) compared to 16.4$\pm$1.0 of the whole comet nucleus (at phase angle 120.2$^\circ$), which could be attributed to a presence of water ice grains amongst the ejecta.

Although the jets were either observed from shadowed area (H-1, H-2 and H-3) or under very low spatial resolution (W-1 at 7.6 m/px, W-2 at 6.0 m/px), pre- and post-perihelion high-resolution images allowed us to observe the morphology of their sources in better details (see Fig.~\ref{fig:jets}). All three Hatmehit jets originated from flat but moderately rough area of the Hatmehit terrains, with jet H-1 closest to the Hatmehit rim ($\sim$23 m) whereas the other two were at close distances to the boulder-rich portion of the depression. The locations of the other two jets feature more diverse morphologies: W-1 came from a fractured ``knobby'' surface and was $\sim$30 m from a higher layer (see Fig.~\ref{fig:Abydos_stretch}) and W-2 was probably located under the ``walls'' that serve as the Wosret/Bastet boundary, on an area covered in talus deposits that was dominated mostly by smaller-sized boulders (0.8-3m, see Figs. 1 and 2 of \citealt{Lucchetti2016}). 

The main activity mechanism of the Abydos surroundings appears to be local insolation, as the jet sources were either submerged in shadows completely at their time of observation (i.e. the Hatmehit jets) or under/close to the shadows cast by nearby structures, similar to the majority of active jets during the perihelion passage \citep{Fornasier2019a}. The fractures of the Bastet region (see Fig.~\ref{fig:Abydos_stretch}) could have played a role in producing jet W-1, as they permitted the heat wave to propagate through underlaying volatile-rich strata \citep{Belton2010, BruckSyal2013}. The mini-outburst W-2 were probably triggered by different mechanisms, one being a reservoir of volatiles below the comet surface as suggested for an outburst in the Imhotep region in July 2016 \citep{Agarwal2017}. Cliff collapse \citep{Vincent2016, Pajola2017} is also a possible explanation, as the W-2 source was probably located under the Wosret/Bastet boundary ``walls'' with relatively high value of local gravitational slope (between approximately 20$^\circ$ and 50$^\circ$, \citealt{Lucchetti2016}), however we did not find an evidence of a morphological change in the area under the boundary.

\section{Conclusion}

The final landing site of Philae shows a number of similarities to the nucleus of comet 67P on a whole. It was located on a dark and moderately red terrain that has been linked with mixtures containing organics \citep{Filacchione2016, Quirico2016}. As observed for the whole nucleus, Abydos and the surrounding terrains show the spectral phase reddening phenomenon i.e. spectral slope increasing with phase angle, varying over time. The spectral reddening coefficients evolved and decreased close to perihelion as observed elsewhere, partially due to the removal of the dust coating. The linear coefficents of phase reddening calculated for Abydos are lower than what were reported in the northern hemisphere of comet 67P \citep{Fornasier2015} and other local regions \citep{Feller2016}. Phase reddening is a common behaviour observed in many Solar System objects, including asteroids 110 Lydia \citep{Taylor1971}, 433 Eros \citep{Clark2002}, 21 Lutetia \citep{Magrin2012}, the Moon \citep{Gehrels1964}, Mercury \citep{Warell2008} and the three Uranian moons Ariel, Titania and Oberon \citep{Nelson1987}; and this phenomenon has been attributed to multiple scattering at high phase angles and/or small scale surface rougness. Other spectral behaviours also exist such as the phase bluing of asteroid 44 Nysa \citep{Rosenbush2009} or the arched shape of the spectral slope of the Martian surface as a function of the phase angle \citep{Guinness1981}. By combining numerical simulation and laboratory experiments, \citealt{Schroder2014} showed that smooth surfaces resulted in an arched shape, whereas microscopically rough regolith results in a monotonous phase reddening. On the other hand, \citealt{Grynko2008} studied the effects of particle size and scattering on phase reddening, and they found that single particle scattering produces monotonous phase dependence on the spectral slope while scattering by multiple components results in a non-monotonous dependence. The same study also show that phase reddening also depends on particle size, as particles larger than 250 microns\footnote{The study was conducted upon a color ratio of 2.4$\mu$m/1.2$\mu$m.} result in phase bluing instead of reddening. Therefore, the phase reddening of comet 67P is probably caused by a microscopically rough regolith that covers the surface of the comet, and the relatively low coefficients of Abydos may indicate that its regolith layer is thinner than in other regions.

Unlike the spectral slope, the reflectance of Abydos did not appear to experience seasonal variations or any clear evolution trend. One possible explanation is that while the site likely experienced dust removal during the perihelion passage that exposed more volatile-rich contents below the surface, these volatiles were embedded below the outer layers or mixed with the dark terrains of the comet, making the dark terrain the dominant optical medium. Other possible factors include non-optimal observing conditions (especially before perihelion) and different spatial resolutions between observing sequences (from $\sim$6 cm/px near the end of the Rosetta mission to over 6 m/px during the 2015 perihelion passage).

Bright patches of exposed volatiles were occasionally observed near Abydos throughout the Rosetta mission, but with higher frequency after perihelion, especially alongside the rim that serves as the Hatmehit/Wosret boundary. The spots were typically only a few m$^2$ or smaller, and even the biggest spots were small compared to others found in other regions e.g. the two $\sim$1500 m$^2$ spots in the Anhur region in the big lobe in April and May 2015 \citep{Fornasier2016}. Estimations of local water ice abundance ranges from a few percent to $\sim$50\%, and up to $\sim$80\% in one case, which possibly corresponded to a fresh exposure of volatiles. Many spots were found under the shadows of nearby structures (e.g. boulders, terraces), and the longest-lived spots (i.e. up to a few months) were found at a relatively colder location. Frosts were sometimes observable near Abydos but only after perihelion, which was also a common behaviour observed in other regions of comet 67P like Anhur \citep{Fornasier2019b}, with the notable exception of Hapi were frost were observed pre-perihelion \citep{DeSanctis2015}.

\section*{Acknowledgements}

OSIRIS was built by a consortium led by the Max-Planck-Institut f\"ur Sonnensystemforschung, Goettingen, Germany, in collaboration with CISAS, University of Padova, Italy, the Laboratoire d'Astrophysique de Marseille, France, the Instituto de Astrof\'isica de Andalucia, CSIC, Granada, Spain, the Scientific Support Office of the European Space Agency, Noordwijk, The Netherlands, the Instituto Nacional de T\'ecnica Aeroespacial, Madrid, Spain, the Universidad Polit\'echnica de Madrid, Spain, the Department of Physics and Astronomy of Uppsala University, Sweden, and the Institut f\"ur Datentechnik und Kommunikationsnetze der Technischen Universitat Braunschweig, Germany. \\
The support of the national funding agencies of Germany (DLR), France (CNES), Italy (ASI), Spain (MEC), Sweden (SNSB), and the ESA Technical Directorate is gratefully acknowledged. We thank the Rosetta Science Ground Segment at ESAC, the Rosetta Mission Operations Centre at ESOC and the Rosetta Project at ESTEC for their outstanding work enabling the science return of the Rosetta Mission. We acknowledge the financial support from the France Agence Nationale de la Recherche (programme Classy, ANR-17-CE31-0004). VH, SF, PHH and MAB acknowlege founding support from CNES.

\section*{Data availability}
The data underlying this article are available in \url{https://www.cosmos.esa.int/web/psa/rosetta}.




\bibliographystyle{mnras}
\bibliography{Abydos_bibliography} 



\appendix

\section{Additional figures}

\begin{figure}
	\includegraphics[width=\columnwidth]{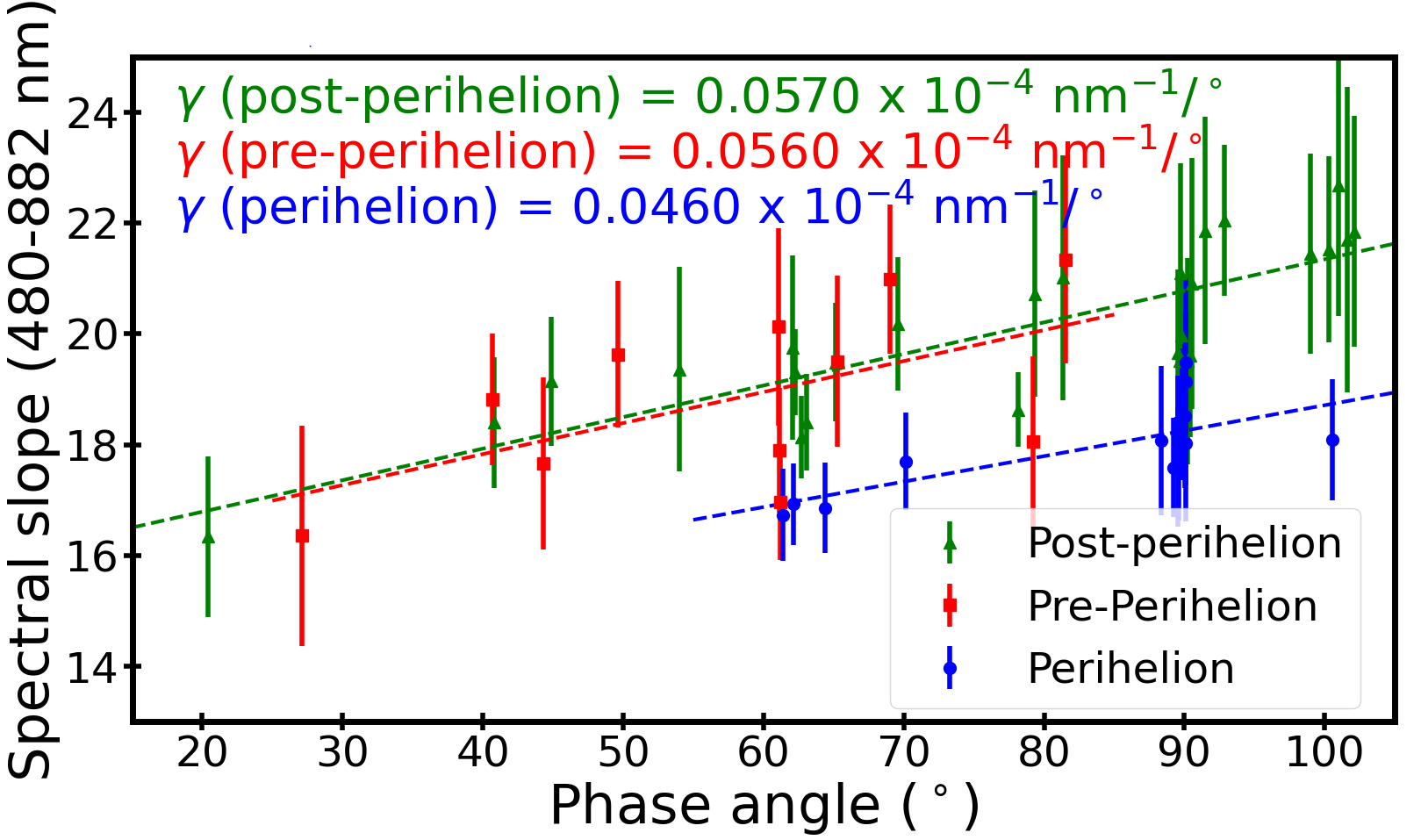}
    \caption{The spectral slope in the 480.7-882.1 nm wavelength range. ``Perihelion" is defined the same way as in Fig.~\ref{fig:Slope}.}
    \label{fig:Slope2441}
\end{figure}

\begin{figure*}
\centering
    \subfloat[][18/02/2015, 13h05 \newline($\alpha= 84.2^\circ$)]{\includegraphics[width=0.22\textwidth]{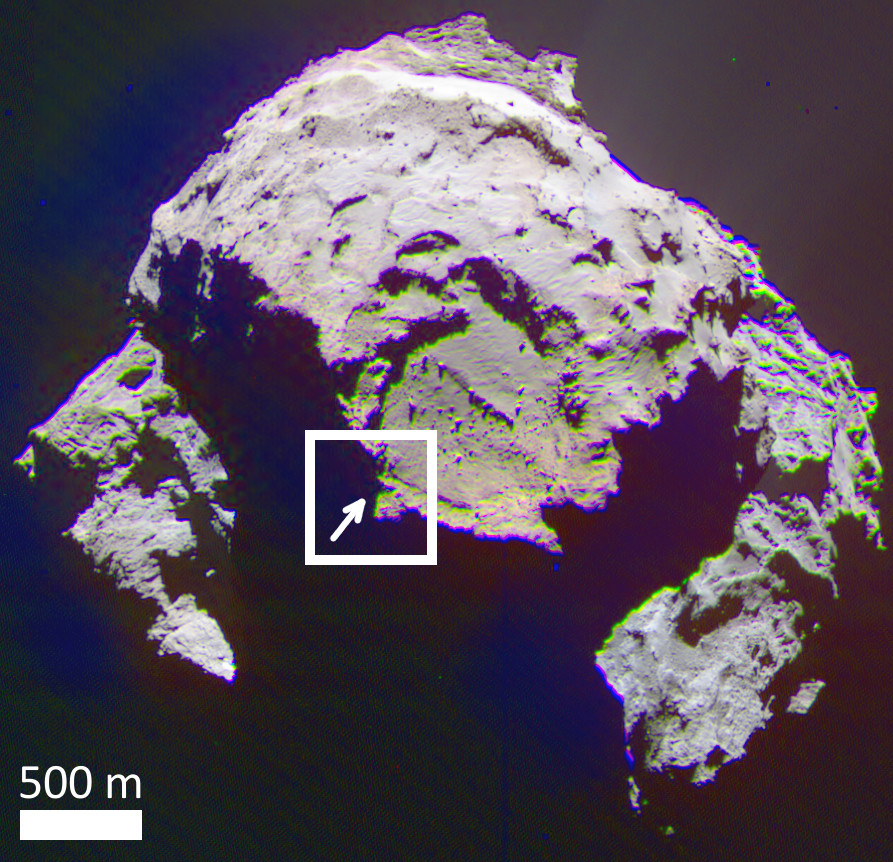}\label{fig:Slope90_a}}
\quad
    \subfloat[][27/06/2015, 7h15 \newline($\alpha= 90.0^\circ$)]{\includegraphics[width=0.22\textwidth]{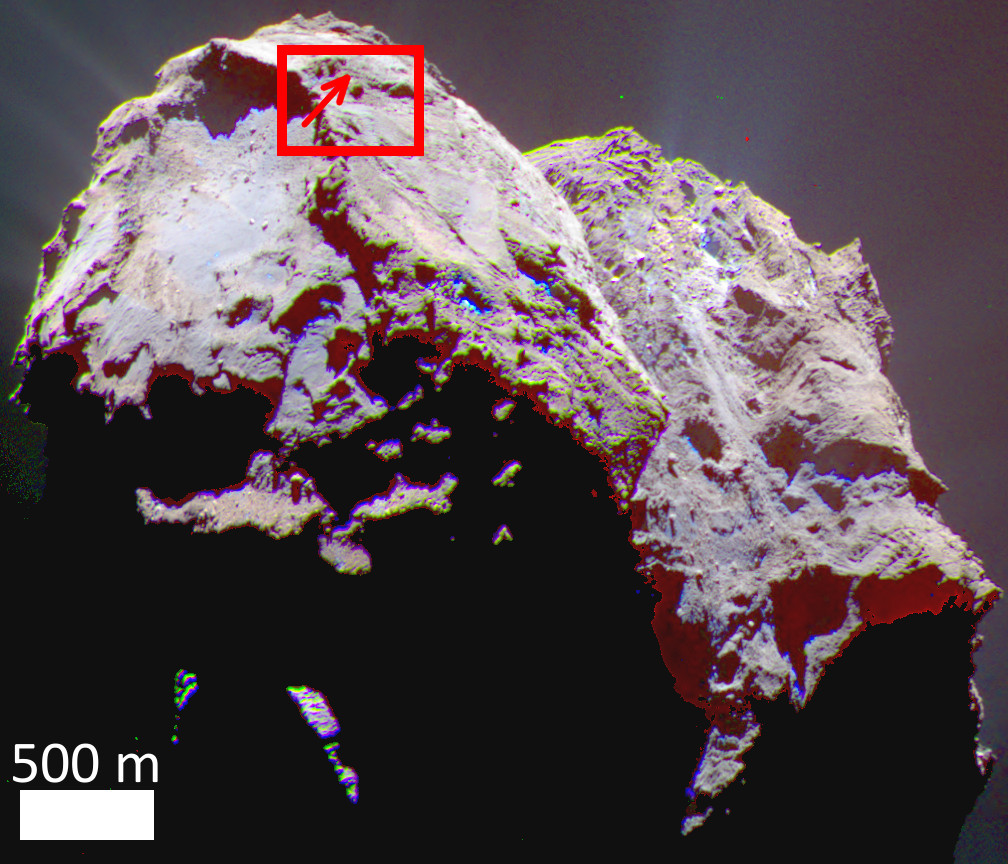}\label{fig:Slope90_b}}
W\quad
    \subfloat[][09/01/2016, 16h06 \newline($\alpha= 90.5^\circ$)]{\includegraphics[width=0.22\textwidth]{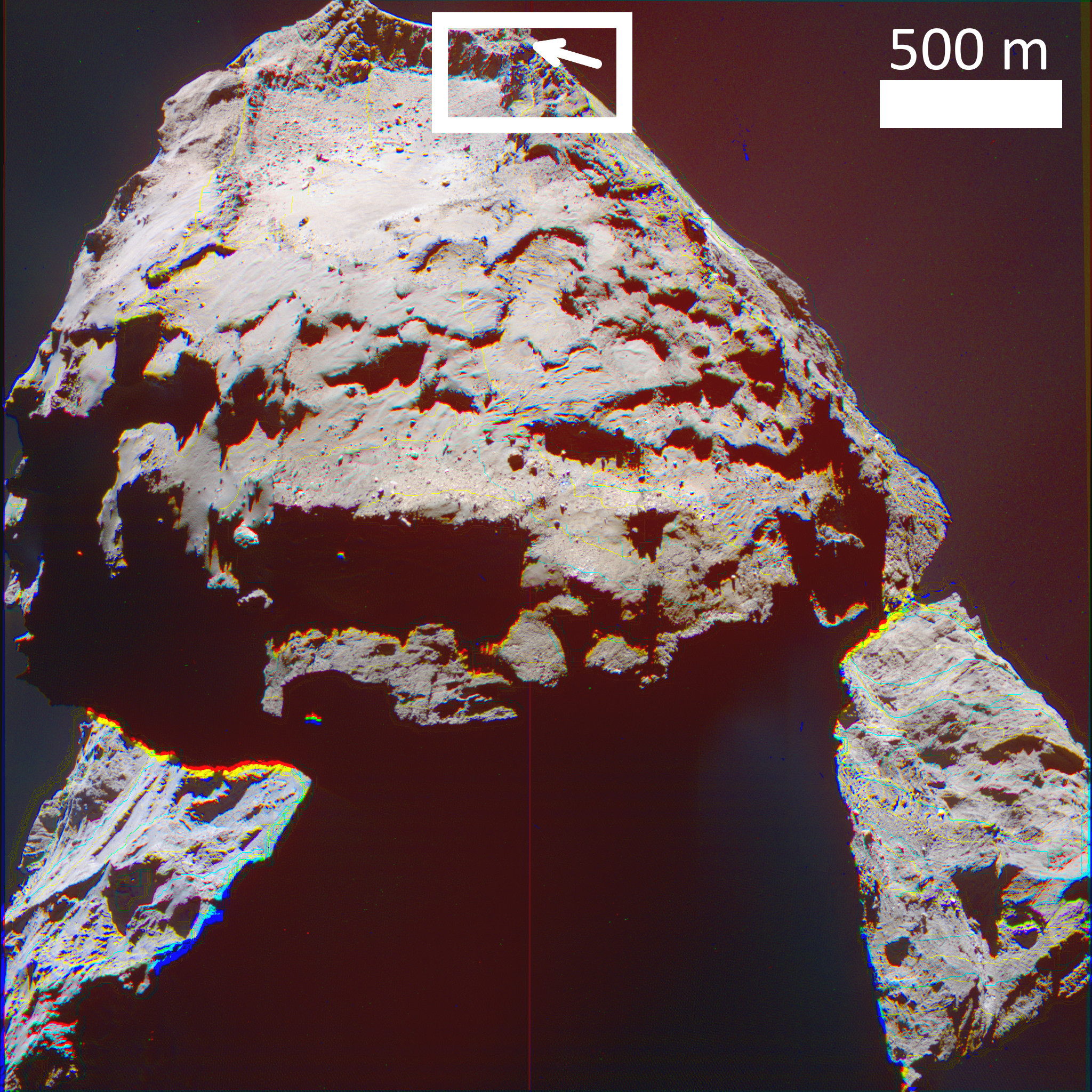}\label{fig:Slope90_c}}
\quad
    \subfloat[][02/07/2016, 15h29 \newline($\alpha= 92.9^\circ$)]{\includegraphics[width=0.22\textwidth]{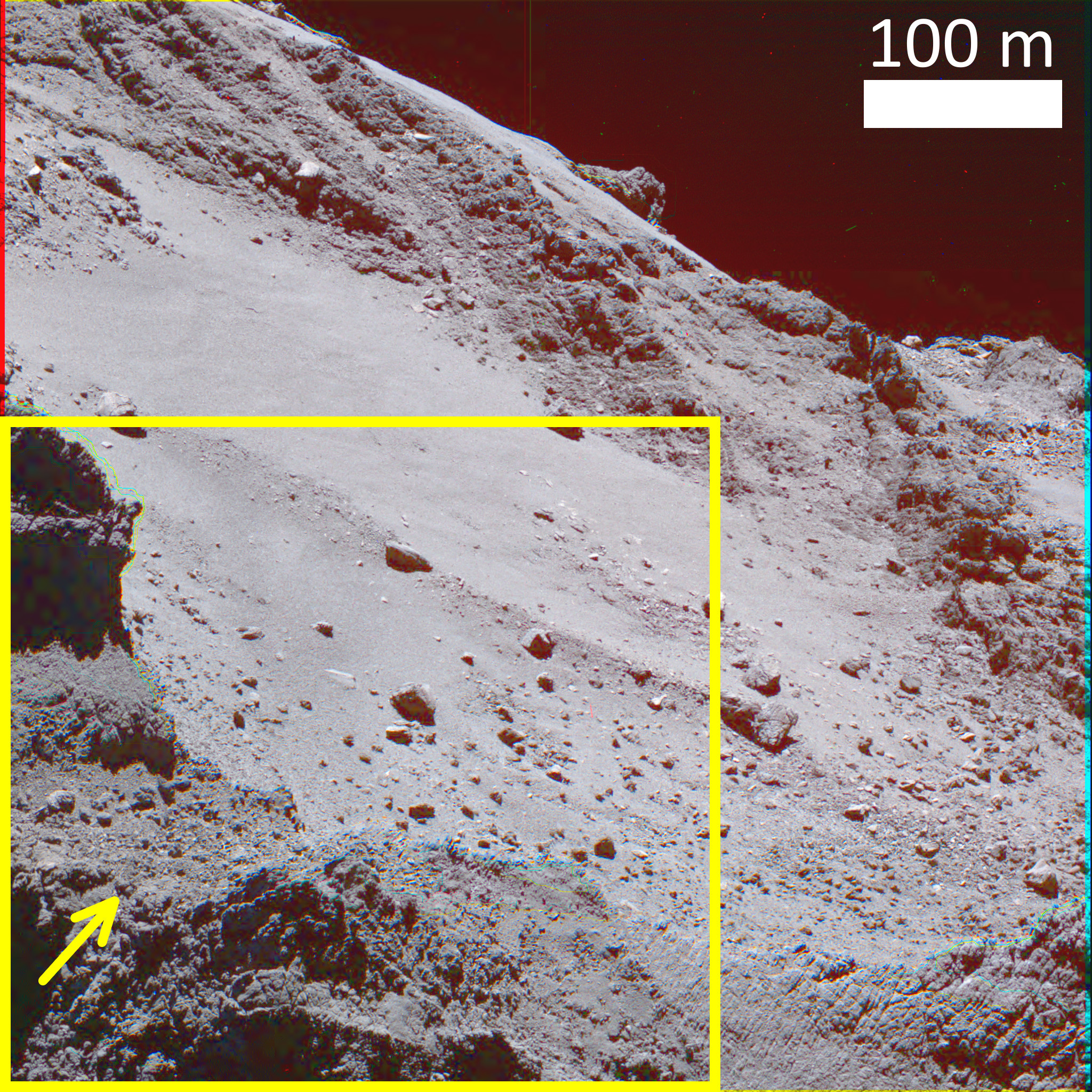}\label{fig:Slope90_d}}
\newline
    \subfloat[][slope= 19.8$\pm$1.7 \%/(100 nm)]{\includegraphics[width=0.22\textwidth]{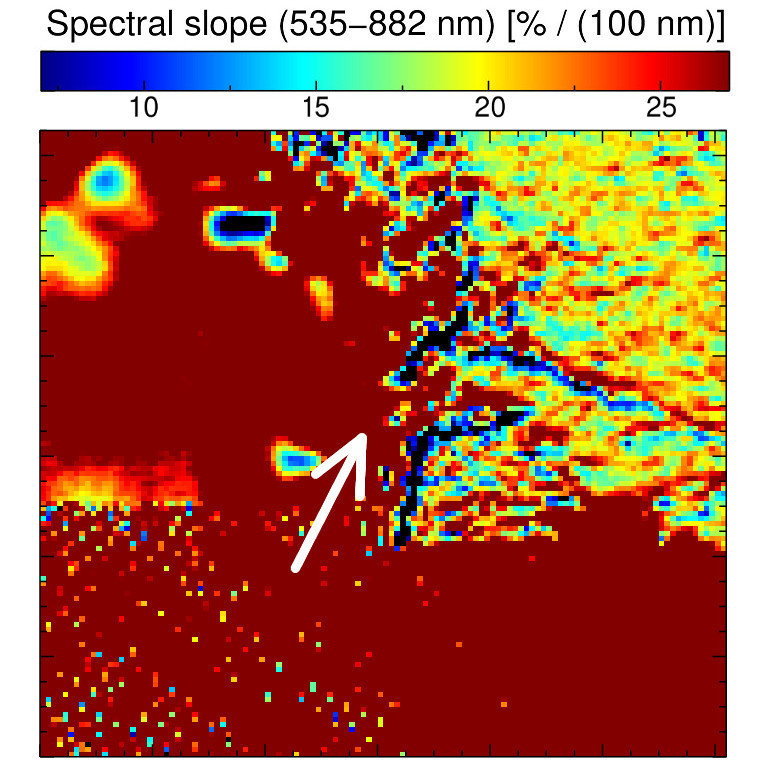}\label{fig:Slope90_e}}
\quad
    \subfloat[][slope= 16.7$\pm$1.0 \%/(100 nm)]{\includegraphics[width=0.22\textwidth]{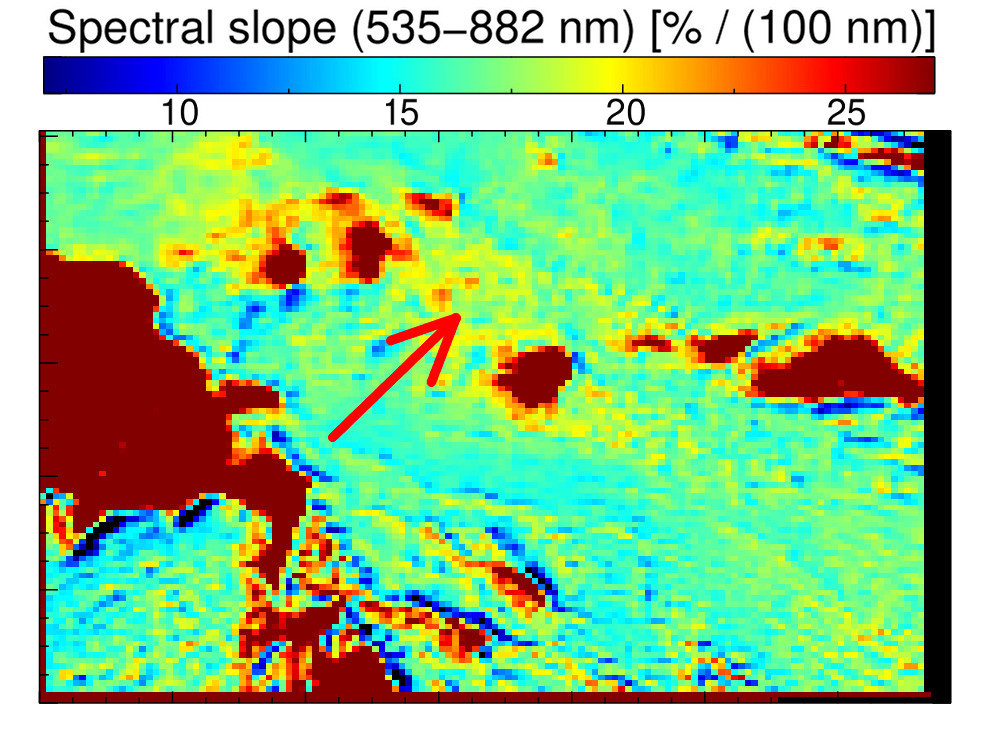}\label{fig:Slope90_f}}
\quad
    \subfloat[][slope= 18.7$\pm$2.1 \%/(100 nm)]{\includegraphics[width=0.22\textwidth]{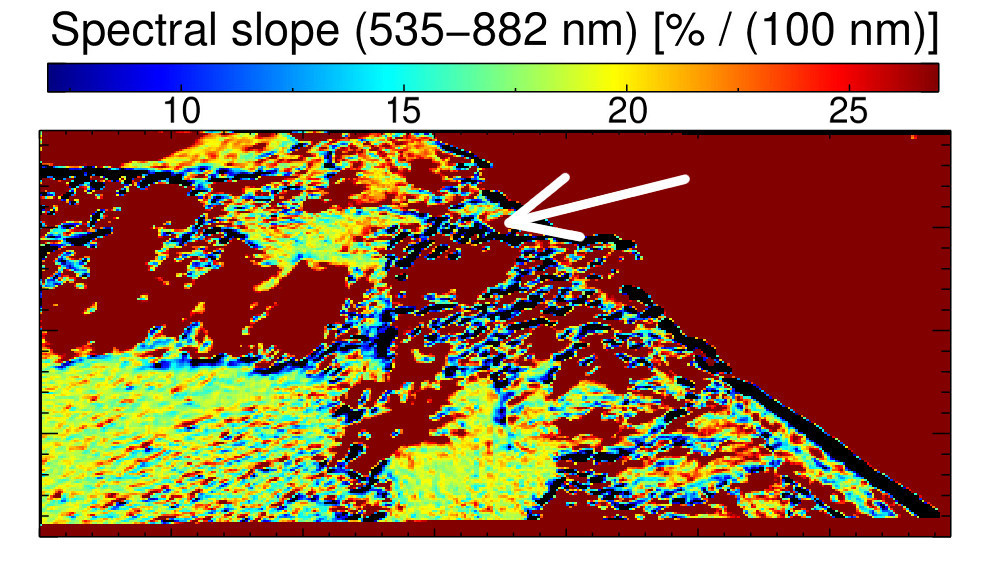}\label{fig:Slope90_g}}
\quad
    \subfloat[][slope= 19.6$\pm$1.7 \%/(100 nm)]{\includegraphics[width=0.22\textwidth]{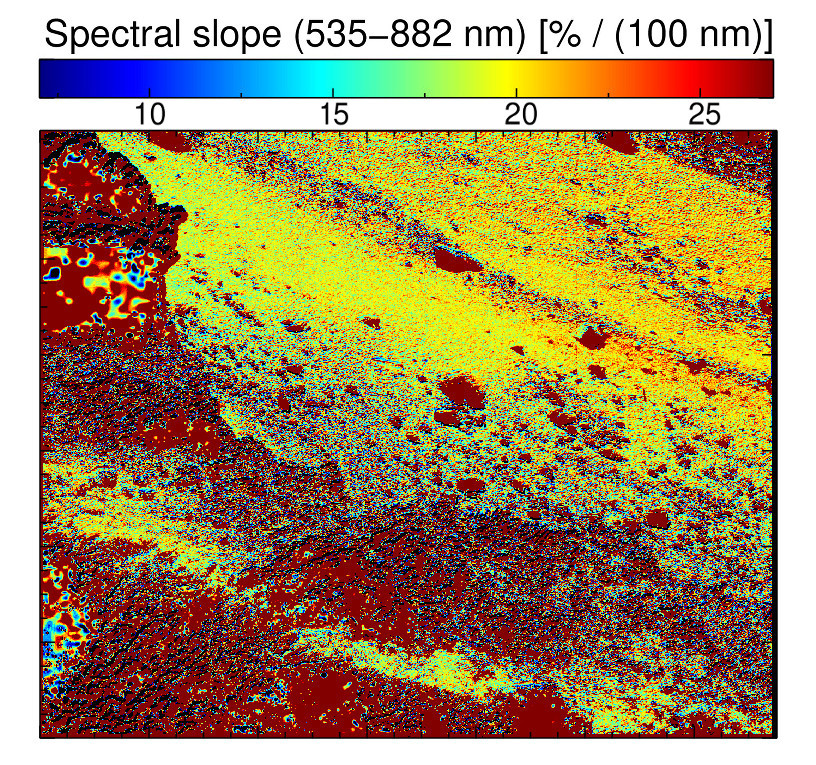}\label{fig:Slope90_h}}
\caption{\textit{Top}: The comet nucleus as imaged by NAC at phase angles $\alpha \sim 90^\circ$ on four different days, in which each box indicates the area around Abydos. \textit{Bottom}: Spectral slope maps (535-882 nm) of the area within the corresponding boxes, where reddest areas are often artefacts related to shadowed regions. Abydos is indicated by an arrow in every panel of this figure.}
\label{fig:Slope90}
\end{figure*}

\begin{figure*}
\centering
    \subfloat{\includegraphics[width=0.2\textwidth]{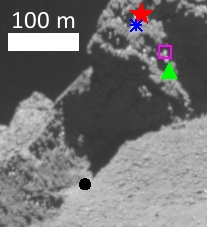}\label{fig:Spectra_Dec25_1}}
\quad
    \subfloat{\includegraphics[width=0.35\textwidth]{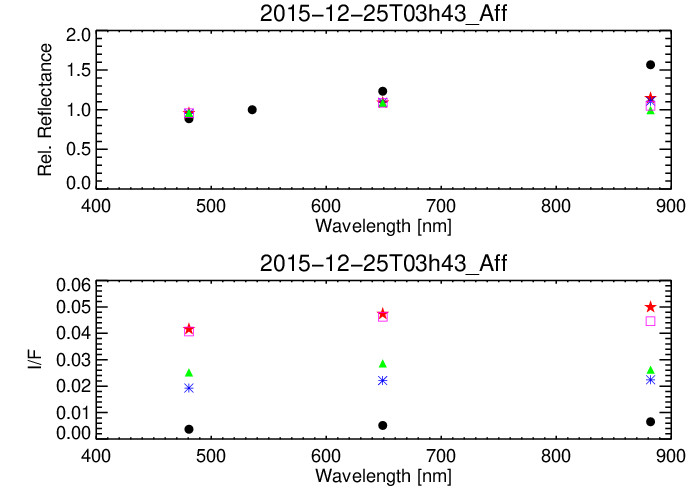}\label{fig:Spectra_Dec25_2}}
\quad
    \subfloat{\includegraphics[width=0.35\textwidth]{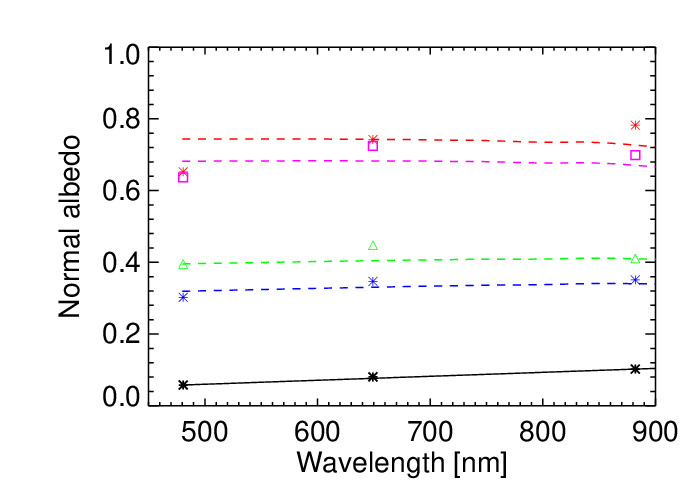}\label{fig:Spectra_Dec25_3}}
\caption{\textit{Left}: The Hatmehit rim near Abydos as imaged by NAC/F22 on 25 December 2015, 3h44. \textit{Middle}: Spectra of several chosen points on the left. Note that the time shown on top of the plot is the timestamp of the image, which is about 70 seconds earlier than the start of the acquisition time. \textit{Right}: The reflectances of the chosen points after correction to zero phase angle. The black line is a linear fit of the dark terrain while the red and blue dotted lines are the best fits of the compositional model for 30$\mu$m ice grain size, indicating water ice abundances of 72.9\% (red), 27.8\% (blue), 35.9\% (green) and 66.3\% (magenta).}
\label{fig:Spectra_Dec25}
\end{figure*}

\begin{figure*}
\centering
    \subfloat[][07/06/2016, 11h50]{\includegraphics[width=0.3\textwidth]{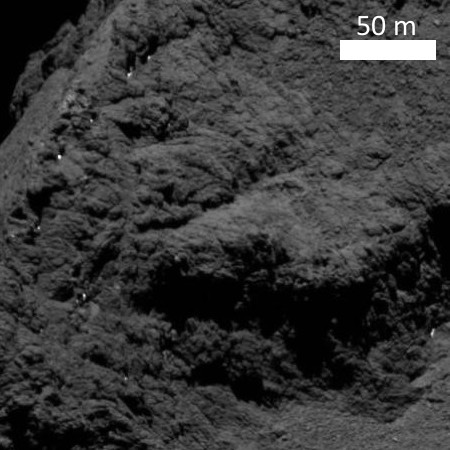}\label{fig:spots_Hatmehit_late_1}}
\quad
    \subfloat[][18/06/2016, 12h21]{\includegraphics[width=0.3\textwidth]{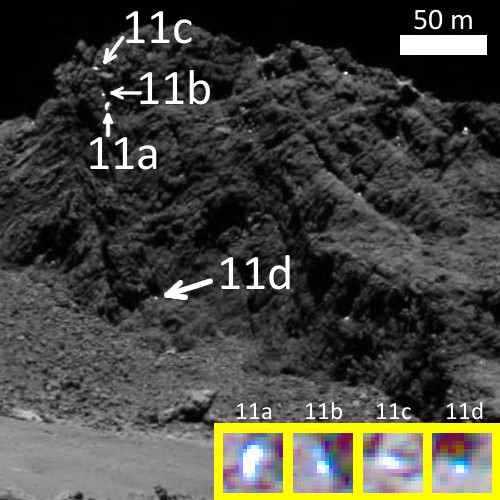}\label{fig:spots_Hatmehit_late_2}}
\quad
    \subfloat[][30/09/2016, 1h21]{\includegraphics[width=0.3\textwidth]{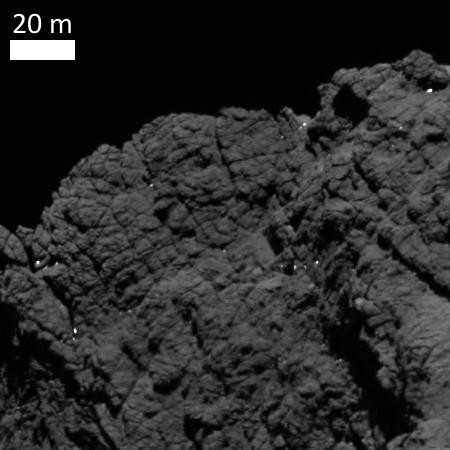}\label{fig:spots_Hatmehit_late_3}}
\caption{Parts of the Hatmehit rim as imaged by NAC/F22 near the end of the Rosetta mission. The middle panel is superposed with 4$\times$ magnifications of spots 11a-d.}
\label{fig:spots_Hatmehit_late}
\end{figure*}

\begin{figure*}
\centering
    \subfloat{\includegraphics[width=0.28\textwidth]{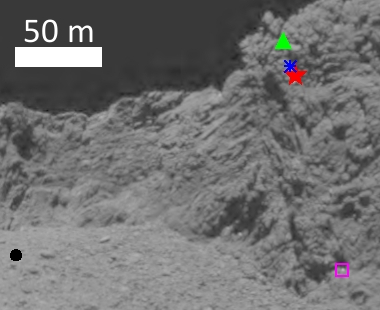}\label{fig:Spectra_18Jun16_1}}
\quad
    \subfloat{\includegraphics[width=0.32\textwidth]{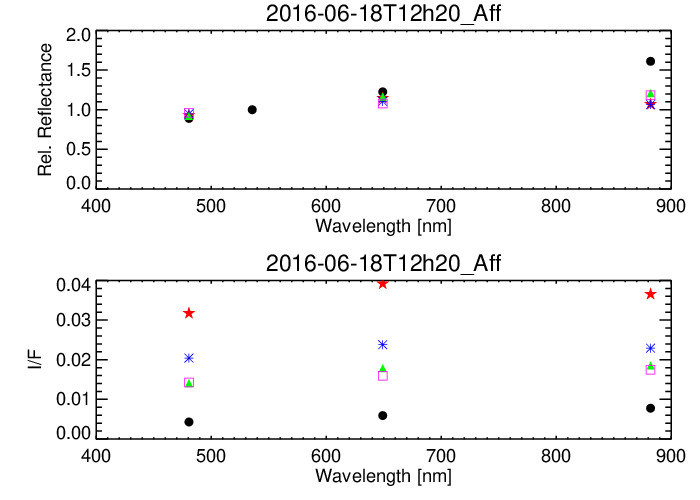}\label{fig:Spectra_18Jun16_2}}
\quad
    \subfloat{\includegraphics[width=0.32\textwidth]{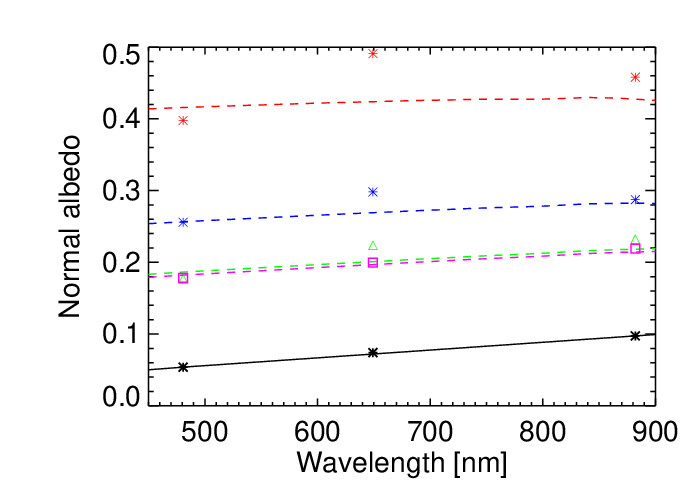}\label{fig:Spectra_18Jun16_3}}
\caption{\textit{Left}: Parts of the Hatmehit rim as captured by NAC/F22 on 18 June 2016, 12h21. \textit{Middle}: Spectra of the chosen points compared to a reference terrain (black), with spots 11a-d respectively represented by the red star, blue asterisk, green triangle and magenta square. Note that the time shown on top of the plot is the timestamp of the image, which is about 70 seconds earlier than the start of the acquisition time. \textit{Right}: The reflectances of the chosen points after correction to zero phase angle. The black line is a linear fit of the dark terrain while the dotted lines are the best fits of the compositional model for 30$\mu$m ice grain size, indicating the water ice abundances of 39.7\% (red), 21.8\% (blue), 14.6 \% (green) and 13.7\% (magenta).}
\label{fig:Spectra_18Jun16}
\end{figure*}


\bsp	
\label{lastpage}
\end{document}